\documentclass[12pt]{article}

\usepackage{graphicx}
\usepackage{amsmath}
\usepackage{amssymb}
\usepackage[british]{babel}
\usepackage{slashed}
\usepackage{afterpage}
\include{epsf}

\def\appendix#1{\addtocounter{section}{1}\setcounter{equation}{0}
\renewcommand{\thesection}{\Alph{section}}
\section*{%Appendix~
\thesection\protect\indent \parbox[t]{11.715cm} {#1}}
\addcontentsline{toc}{section}{Appendix\thesection\ \ \ #1} }
\newcommand{\id}{{1\!\!1}} %% identity operator

\def\bra#1{\left\langle #1\right|}
\def\ket#1{\left| #1\right\rangle}
\def\hs#1#2{\left\langle #1\right|\left. #2\right\rangle}

\def\be{\begin{equation}}
\def\ee{\end{equation}}
\def\bea{\begin{eqnarray}}
\def\eea{\end{eqnarray}}

\newcommand{\ii}{{\mathrm i}}

\newcommand{\eqn}[1]{(\ref{#1})}

\begin{document}

\begin{titlepage}
\begin{flushright}
ICCUB-12-312
\end{flushright}

\begin{center}

\baselineskip=24pt

%{\Large\bf Numerical analysis\\
% of  scalar field theory on the fuzzy disc}

{\Large\bf Noncommutative Field Theory:\\
 Numerical Analysis with  the Fuzzy Disc}

\baselineskip=14pt

\vspace{1cm}

{Fedele Lizzi$^{1,2,3}$ and Bernardino Spisso$^1$}
\\[6mm]
$^1${\it Dipartimento di Scienze Fisiche, Universit\`{a} di Napoli
{\sl Federico II}}
\\[4mm]
$^2${\it INFN, Sezione di Napoli}\\
{\it Monte S.~Angelo, Via Cintia, 80126 Napoli, Italy}
\\[4mm]
$^3$ {\it Departament de Estructura i Constituents de la Mat\`eria,
\\Institut de Ci\'encies del Cosmos,
Universitat de Barcelona,\\
Barcelona, Catalonia, Spain}
\\{\small\tt
fedele.lizzi@na.infn.it, nispisso@tin.it}
%\\[10mm]

\end{center}

\vskip 2 cm

\begin{abstract}
The fuzzy disc is a discretization of the algebra of functions on
the two dimensional disc using finite matrices which preserves the
action of the rotation group. We define a $\varphi^4$ scalar field
theory on it and analyze numerically for three different limits for
the rank of the matrix going to infinity. The numerical simulations
reveal three different phases: uniform and disordered phases already
the present in the commutative scalar field theory and a nonuniform
ordered phase as a noncommutative effects. We have computed the
transition curves between phases and their scaling. This is in
agreement with studies on the fuzzy sphere, although the speed of
convergence for the disc seems to be better. We have performed also
three the limits for the theory in the cases of the theory going to
the commutative plane or commutative disc. In this case the theory
behaves differently, showing the intimate relationship between the
nonuniform phase and noncommutative geometry.
\end{abstract}

\end{titlepage}

\section{Introduction}

Field theory on noncommutative spaces (for a review
see~\cite{Szaboreview}) are a interesting arena to study the
properties of noncommutative geometry~\cite{Connesbook, Landi,
Ticos}. We will not review here the motivations for a noncommutative
geometry. Among the applications of noncommutative geometry are
fuzzy spaces. The original fuzzy space is the fuzzy
sphere~\cite{Fuzzy-s}, but other spaces have been built  (for a
review see~\cite{BalSeckinSachin}). Field theory on fuzzy spaces are
a way to study field theory on noncommutative spaces on a finite
setting. The fact that the approximation is based on matrices makes
them ideal for numerical studies. The most studied case is the fuzzy
sphere, for a review see~\cite{Medina,fuzzy-numer-3}. Most of the interest for
these investigation is the presence of the limit for which the rank
of the matrices grow, and at the same time the radius of the sphere
increases, thus recovering in the limit a noncommutative plane. Here
and in the following by noncommutative plane we mean the algebra
generated by the noncommuting variables $x$ and $y$ with commutator
\be
[x,y]=\ii\theta \label{commrel}
\ee
This sort of plane is usually described by a noncommutative $*$
product among the fields, such as the Gr\"onewold-Moyal
product~\cite{Gronewold, Moyal}. Other product are possible, and a
comparison among them is given in~\cite{GLV08}. For our purposes it
is however better to think of the ``quantization'' of the filed as a
map from functions $f$ to operators $\hat f$. As we shall see the
fuzzy disc approximation will render the matrices finite, and hence
amenable to numerical simulations.

We will study then a quantized $\varphi^4$ scalar field theory
approximating field with $N\times N$ matrices, and perform three
different limits, giving rise to the noncommutative plane, the
commutative plane and the commutative disc. We are interested in
particular to the phase transitions \cite{montecarlo-3,montecarlo-4} of the theory as we change the
parameters of the action. The quantity of interest are
susceptibility and specific heat, as well as other order parameters,
which we will describe below.

Is worthwhile compare the present method with  the simulations of such theories on the lattice. In general the simulations of scalar theories on fuzzy spaces are
slower than in their lattice counterparts since the fuzzy models  due to the self-interaction term $\varphi^4$ are intrinsically non local and for higher  power of the self-interacting  the number of operations to calculate each Monte Carlo step $\delta S$ grows even faster.  However,  we can  expect some advantages of the fuzzy approach with the simulations of other field theories, in particular the fermionic theories or super symmetric models \cite{Fuzzy-susy}. Beside another recent field of application the matrix models are the Yang-Mills actions \cite{Fuzzy-YM,Num-WG1,Num-WG2} in which appear the concept of the emerging geometry.

The paper is organized as follows. In the next section will be
introduce the fuzzy disc and its properties. The real scalar field
theory on the fuzzy disc will be presented in
section~\ref{se:scafieldtheo}. The simulation and the result are in
Sect.~\ref{se:simulations}, where also the phase boundaries between
the three phases present and their scaling properties are derived.
Finally, the results are summarized and discussed in the
conclusions.

\renewcommand{\aa}{\mathsf{\hat a}}

\section{The fuzzy disc}
\setcounter{equation}{0}
The construction of the fuzzy disc~\cite{fuzzy1, fuzzybal, fuzzy2, fuzzy3,
fuzzy4} is based on a quantization of the function on the plane,
i.e.\ the association to a function of an operator on an infinite
dimensional Hilbert space $F$. Consider the Hilbert space generated
by the countable basis $\ket{\psi_n}$ with $n$ positive or zero.
Consider the operator $\aa$ and its hermitian conjugate
$\aa^\dagger$ acting on the basis as
\bea
\aa\ket{\psi_n}&=& \sqrt{n\theta}\ket{\psi_{n-1}}, \nonumber \\
\aa^\dagger\ket{\psi_n} & = & \sqrt{(n+1)\theta}\ket{\psi_{n+1}}
\label{aad},
\eea
With the commutation rule:
\be
\left[\aa,\aa^\dagger\right]=\theta \id .
\ee
One recognizes of course the usual creation and annihilation
operator of single harmonic oscillator, with a slightly unusual
normalization (in the following we will revert to the usual
one). In terms of the variable of~\eqn{commrel}.

Consider a function on the plane in terms of the conjugate variables $\bar z, z$ with $z=x+\ii y$
we define a quantization map which associates to a function  an operator  $\hat f$ according to the following rule:
\be
f(z,\bar{z})=\sum_{m,n=0}^\infty f_{mn}^{Tay}\bar{z}^m z^n\Leftrightarrow \hat{f} =\sum_{m,n=0}^\infty
f_{mn}^{Tay}\aa^{\dagger m}\aa^n.  \label{map1}
\ee
Here and in the following we will not discuss the issues of convergence of the series, we assume the coefficients are of rapid decrease. 
A more rigorous construction, based on the coherent states of the Heisenberg group and on a Weyl-Wigner map can be found in~\cite{fuzzy3,Zampthesis}.

This quantization associates an operator to a function, the inverse map is 
\be
f=\langle{z}|\hat f |z\rangle.
\ee
with $|z\rangle$ the usual coherent states $\aa|z\rangle=z|z\rangle$. The function associated to the operator is called the \emph{symbol} of the operator.

More generally we can consider operators written in a density matrix notation:
\be
\hat{f}=\sum_{m,n=0}^\infty f_{mn}\ket{\psi_m}\bra{\psi_n}. \label{densitymatrixbasis}
\ee
To this operator is associated the function:
\be f(z,\bar{z})=e^{-\frac{z \bar{z}}{\theta}}\sum_{m,n=0}^\infty f_{nm}\frac{z^n \bar{z}^m}{\sqrt{m!n!\theta^{n+m}}}. \label{simden}
\ee
Where the relation between the Taylor coefficients $f_{mn}^{Tay}$ and the $f_{mn}$ is
\be
f_{lk}=\sum_{q=0}^{min(l,k)}f_{l-q \ k-q}^{Tay}\frac{\sqrt{k!l!\theta^{k+l}}}{q!\theta^q}.
\ee

After quantization if the plane using the map \eqref{map1}  and its inverse we define the subalgebras of finite $(N+1)\times (N+1)$ matrices considering the functions with a truncated expansion, more precisely taking in account  only the expansion terms if both $n$ and $m$  are smaller than
a given integer $N$. Keeping $\theta$ fixed the limit $N\rightarrow\infty$ the truncated algebra tends to the noncommutative plane. This is obtained with the help of the projection operator
\be
\hat{P}^{(N)}_\theta=\sum_{n=0}^N \ket{\psi_n}\bra{\psi_n}.
\ee
To this operator corresponds the function
\be
P_\theta^{(N)}(\bar{z},z)=\sum_{n=0}^N \hs{z}{\psi_n}\hs{\psi_n}{z}=e^{-\frac{z\bar{z}}{\theta}}\sum_{n=0}^N \frac{(z \bar{z})^n}{n!\theta^n}=e^{-\frac{r^2}{\theta}}\sum_{n=0}^N\frac{r^{2n}}{n!\theta^n}.
\ee
This sum can be expressed in terms of incomplete gamma function and gamma function obtaining a radial function:
\be
P_\theta^{(N)}(r,\varphi)=\frac{\Gamma\left(N+1, r^2 / \theta\right)}{\Gamma\left(N+1 \right)}.\label{gamma projectors}
\ee
In the limit $N \rightarrow\infty$ if $\theta$ is fixed and nonzero the symbol \eqref{gamma projectors} converges pointwisely to the constant function
$P_\theta^{(N)}(r,\varphi)= 1$ and recovering the noncommutative plane. Otherwise if the product $N\theta$ is fixed equals to a real constant $R^2$ the limit for $N \rightarrow\infty$ of the \eqref{gamma projectors} become:
\be
P_\theta^{(N)}(r,\varphi) \rightarrow\left[\begin{array}{cc} 1&r<R \\ 1/2&r=R \\ 0&r>R \end{array}  \right]=\id_d (r).
\ee
Where the convergence still pointwise. This limit converges to a step function in the radial coordinate $r$ in other words the characteristic function of the disc on the plane. Given the algebra ${\mathcal A}$ generated by $\aa$ and $\aa^\dagger$, a sequence of sub algebras $A^{(N)}_\theta$ can be defined by:
\be
\mathcal{A}_\theta^{(N)}=\hat P_N {\mathcal A} \hat P_N.
\ee
 The effect of this projection on a generic function is:
\be
f_\theta^{(N)}=\bra {z}P_\theta^{(N)} \hat{f}P_\theta^{(N)} \ket{z}=e^{-\frac{z \bar{z}}{\theta}}\sum_{m,n=0}^N
f_{nm}\frac{z^n \bar{z}^m}{\sqrt{m!n!\theta^{n+m}}},
\ee
namely a truncation of the series expansion \eqref{simden}. The full algebra $A$ is isomorphic to an algebra of operators however the previous relations shows that $A^{(N)}_\theta$ is isomorphic to $\mathbb{M}_{N+1}$ the algebra of $(N + 1)$ rank matrices, on each sub algebra $A^{(N)}_\theta$ the symbol
\eqref{gamma projectors} is then the identity matrix in every $\mathbb{M}_{N+1}$. The name fuzzy disc  derives from the cutoff; for a fixed $N$ every function with the same first $(N+1)^2$  terms are mapped in the same matrix losing all the informations of the higher orders. In addiction the symbols of the fuzzyfied function are still defined outside the disc of radius $R^2=N\theta$, but they are exponentially damped outside the disc. 

It s convenient to use a related way is to define the fuzzy disc using an association that maps directly  a function defined on the disc of radius $R$ to a $(N+1)^2$ matrix. Is well know that in polar coordinates $z=\rho e^{i\vartheta}$ a base for $f(\rho,\vartheta)$ are the Bessel functions of integer order which are the eigenfunctions of the Laplacian on the disc with Dirichlet boundary conditions. If $f$ is square integrable with respect to the standard measure on the disc $\textrm{d}x = \rho\textrm{d}\rho\textrm{d}\vartheta$ it can be expanded in terms of Bessel
functions:
\be
f(\rho,\vartheta)=\sum_{n=-\infty}^{+\infty}\sum_{k=1}^\infty f^{Bes}_{nk} e^{i n\vartheta} J_{\mid n\mid}\left(\sqrt{\lambda_{\mid n \mid,k} \rho}\right).
\label{bessel}
\ee
There is also a fuzzy version of the angle variable in terms of the phase operator~\cite{KobayashiAsakawa}.

It is possible to define a fuzzy Laplacian the eigenstates of this fuzzy Laplacian will be the analogous of the standard Bessel
in the sense that the fuzzy Bessel are a base for $\mathbb{M}_{N+1}$ and their symbols tends to the standard Bessel for $N\rightarrow\infty$.
The first step is to define the derivations in $A_\theta^N$ using the projectors. In $A$ we have:
\bea
\partial_z f &=&\frac{1}{\theta}\bra{z} \left[\hat{f},\hat{a}^\dagger \right]\ket{z}, \\
\partial_{\bar{z}} f &=&-\frac{1}{\theta}\bra{z} \left[
\hat{f},\hat{a} \right]\ket{z}.
\eea
Therefore we define
\bea
\partial_z f_\theta^{(N)} &\equiv&\frac{1}{\theta}\bra{z} P_\theta^{(N)}\left[P_\theta^{(N)}
\hat{f}P_\theta^{(N)},\hat{a}^\dagger \right]P_\theta^{(N)}\ket{z}, \\
\partial_{\bar{z}} f_\theta^{(N)} &\equiv&-\frac{1}{\theta}\bra{z}P_\theta^{(N)} \left[P_\theta^{(N)}
\hat{f}P_\theta^{(N)},\hat{a} \right]P_\theta^{(N)}\ket{z}.
\eea
These operators satisfy the Leibniz rule and are a derivations on each $A_\theta^{(N)}$. In the same way for the definition of the Laplacian operator we start from the exact expression on $A$:
\be
\nabla^2f(z,\bar{z})=4\partial_z\partial_{\bar{z}}f(z,\bar{z})=\frac{4}{\theta^2}\bra{z}\left[\hat{a},\left[f,\hat{a}^\dagger
\right]\right]\ket{z}. \ee
We define in each $A_\theta^{(N)}$:
\be
\nabla^2f_\theta^{(N)}(z,\bar{z})=\frac{4}{\theta^2}\bra{z}P_\theta^{(N)}\left[\hat{a},\left[P_\theta^{(N)}f
P_\theta^{(N)},\hat{a}^\dagger \right]\right]P_\theta^{(N)}\ket{z},
\ee
and then
\be
\nabla^2\hat{f}=\frac{4}{\theta^2}P_\theta^{(N)}\left[\hat{a},\left[P_\theta^{(N)}f
P_\theta^{(N)},\hat{a}^\dagger \right]\right]P_\theta^{(N)}.
\ee
Can be proved that the eigenvalues of this Laplacian are the solution of this equation:
\be
\sum_{k=1}^{N+1-n}(-1)^{N+1-n+k}
\frac{(N+1)!(N+1-n)!}{k!(n+k)!(N+1-n+k)}\left(\frac{\lambda^{(N)}_{n,k}}{4N}
\right)^k=0. \label{lam}
\ee
There is a sequence of eigenvalues labeled by $n,k$ where $n$ is the order of the correspondent  Bessel function and $k$ indicates the $k$-th zero.
This eigenvalues can be calculated numerically from \eqref{lam} and they appears to converge for $N\rightarrow\infty$ to the spectrum of the standard Laplacian
defined on a disc with Dirichlet homogeneous boundary conditions on the edge. The eigenstates of the standard Laplacian are:
\be
\hat{\Phi}^{(N)}_{n,k}=\left(\frac{\lambda^{(N)}_{n,k}}{4N}
\right)^{n/2}\sum_{a=0}^{N-n}\sqrt{a!(a+n)!}\left[\sum_{s=0}^a\left(-\frac{\lambda^{(N)}_{n,k}}{4N}\right)^{s}
\frac{1}{s!(s+n)!(a-s)!}\right]\ket{\psi_a}\bra{\psi_{a+n}}.\
\ee
Where the eigenstates with the negative $n$ are hermitian conjugate of $\hat{\Phi}^{(N)}_{n,k}$.  In the fuzzy approximation the continuum eigenfunctions
$ e^{i n\varphi} J_{\mid n\mid}\left(\sqrt{\lambda_{\mid n \mid ,k} r }\right)$  are represented by matrices $M_{N+1}$ but the max possible  $n$ are
fixed by the dimension of the fuzzyfication  $n \leq N$.

\section{Real scalar field theory on the fuzzy disc \label{se:scafieldtheo}}
\setcounter{equation}{0}

The previous construction allows us to define a fuzzyfied version of a field theory on a disc. For the real scalar case with Euclidean signature, with the $\varphi^4$ potential, the action  is given by:
\be
S(\varphi) = \int (\varphi\nabla^2\varphi + \mu\varphi^2 + \frac{\lambda}{4}\varphi^4 )  \textrm{d}^2z, \label{S-com}
\ee
with  $\nabla$ the Laplacian on the disc, $\mu$ and $\lambda$ are respectively a mass and  an interaction parameters. We choose this particular model  because for the plane case it is  well studied. It is known \cite{UV-IR,UV-IR-1} that the diagrammatic expansion of this theory has only one divergent diagram, the so called tadpole diagram and  it is renormalizable. Using the fuzzyfication procedure the action \eqref{S-com} can be approximated by:
\be
\hat{S}_N(\hat{\varphi})=\pi\theta \operatorname{Tr} \left(\hat{\varphi}\hat{\nabla}^2 \hat{\varphi} + \mu\hat{\varphi}^2 +  \frac{\lambda}{4}\hat{\varphi}^4 \right)=\pi\theta \operatorname{Tr} \left(\frac{4}{\theta^2}\hat{\varphi}\left[\hat{a},\left[\hat{\varphi}
,\hat{a}^\dagger \right]\right]+\mu\hat{\varphi}^2 + \frac{\lambda}{4}\hat{\varphi}^4 \right).
\ee
In this action  $\hat\varphi$ are finite dimensional hermitian matrices
and the product between the field became the standard matrix
multiplication,  being a finite matrix model   can be approached
numerically using  Monte Carlo techniques a method currently in use
for other fuzzy spaces like the sphere. Note that the fuzzy approximation provides both an infrared and an ultraviolet regularization of the theory.

We can rescale  the operators $\aa,\aa^\dagger$ \eqref{aad} as $a=\sqrt{\theta}\aa,a^\dagger=\sqrt{\theta}\aa^{\dagger} $  in order the extract the $\theta$ dependence and recast the action making $\theta$ manifest:  
\be
\hat{S}_N(\hat{\varphi})=\pi\operatorname{Tr} \left\{4\hat{\varphi}\left[\hat{a},\left[\hat{\varphi}
,\hat{a}^{\dagger} \right]\right]+\theta\left(\mu\hat{\varphi}^2 + \frac{\lambda}{4}\hat{\varphi}^4 \right)\right\}.\label{S-Ncom}
\ee
It will be useful later  to define the potential part of the action given by (from now to simplify notations  we will omit the hats and the subscript $N$, as all fields are matrices):
\be
S_V(\varphi) = \pi\theta\operatorname{Tr}\left(\mu \varphi^2 + \frac{\lambda}{4}\varphi^4 \right)\label{par-1}.
\ee
In the simulation we will use another equivalent choice of the parameters often used for the potential part namely:
\be
S^\prime_V(\varphi) = \pi\theta r\operatorname{Tr}\left(\mu^\prime\varphi^2 + \frac{\varphi^4}{4} \right)\label{par-2}.
\ee
The two choices are related by
\bea
r&=&\lambda,\nonumber\\
r\mu'&=&\mu.
\eea

Since the scalar field on either
commutative or fuzzy disc  is defined on a finite volume it is impossible to 
have a phase transition, however phase transitions may be found
when the matrix dimension or the radius of the sphere become
infinite.
As already states the fuzzy disc  arise  from the relation $R^2=\theta N$ where $\theta$  is the parameter of non-commutativity and $R$ is the radius of the disc. Introducing in the non-commutativity parameter a dependence  on the matrix size $N$ and  performing  different limits we have  three different cases:

\newcommand{\CD}{{\bf CD}}
\newcommand{\NCP}{{\bf NCP}}
\newcommand{\CP}{{\bf CP}}

\begin{itemize}

\item[\CD]  $N\to\infty$ with $R^2=N\theta$ fixed. In this case we recover the commutative disc of radius $R$. In the following we will take $R=1$. 

\item[\NCP]  $N\to\infty$ with $R\to\infty$ and $\theta$ fixed. In this case we recover the noncommutative plane. We will take $\theta=1$ and  therefore $R^2=N$. 

\item[\CP]  $N\to\infty$, $\theta\to 0$ and $R\to\infty$. In this case we recover the commutative plane. In the following we will the $\theta=1/\sqrt{N}$ which implies $R^2=\sqrt{N}$, in the limit $N \to \infty$ we recover the  commutative plane\footnote{It is possible recover  the commutative plane taking $\theta=N^\alpha$, where $\alpha $ is a real number varying between $ -1<\alpha<0 $, our choice is just dictated by simplicity. }.

\end{itemize}

In the simulation we expect at least one critical line corresponding to the critical behavior of the continuous scalar field on the plane. Indeed one of the main aim
of the simulation will be to find the phase diagrams for the fuzzy field theory.

\section{Simulations \label{se:simulations}}
\setcounter{equation}{0}

Using Monte Carlo method we will produce a sequence of configurations 
$\{\psi_j \}_{j = 1, 2,\cdots,T_{MC}}$ and
evaluate the average of the observables over the set of configurations. These sequences of configurations, called Monte Carlo chain, are representatives of the configuration space at given parameters.  In this framework the expectation value is approximated as
\be
\langle O\rangle \approx \frac{1}{T_{MC}}\sum_{j=1}^{T_{MC}}O_j\; \label{expet},
\ee
where $O_j$ is the value of the observable $O$ evaluated in the $j$-sampled configuration, $\psi_j$, $O_j= O[\psi_j]$.
The internal energy is defined as
\be
E(N,\mu,\lambda)= \langle S\rangle\;,
\ee
It is also very useful to compute separately the average values of the two contributions:
\begin{align}
D(N,\mu,\lambda)&= \langle\operatorname{Tr}\left(S_{D}\right) \rangle \;,
\\
V(N,\mu,\lambda) &= \langle \operatorname{Tr} \left( S_{V}\right)  \rangle \;.
\end{align}
Where $S_{D}$ is defined as:
\be
S_D(\varphi) =4\pi\operatorname{Tr} \left\{\varphi\left[a,\left[\varphi
,a^{\dagger} \right]\right]\right\}.
\ee
The specific heat takes the form:
\be
C(N,\mu,\lambda)= \langle S^2\rangle - \langle S\rangle^2\;.
\ee
These quantities correspond to the usual definitions for energy
\be
E(N,\mu,\lambda) =  -\frac{1}{\mathcal{Z}}\frac{\partial\mathcal{Z}}{\partial\beta}
\ee
and specific heat
\begin{equation}
C(N,\mu,\lambda) =\frac{\partial E}{\partial\beta}\;,
\end{equation}
where $\mathcal{Z}$ is the partition function. Another important observable used to look for a phase transition is   finite-volume susceptibility defined as:
\be
\chi = \langle \operatorname{Tr}(\varphi)^2\rangle - \langle|\operatorname{Tr}(\varphi)|\rangle^2.  \label{sus}
\ee
The phase transitions are  located at the values the previous quantities show a peak when varying the parameters. Note that in comparing the \eqref{sus}  with  the usual susceptibility the second  term is contains  $\langle|\operatorname{Tr}(\varphi)|\rangle^2$, otherwise due to the fact that symmetry causes $\operatorname{Tr}(\varphi) = 0$ there will be no peak even in presence of a phase transition.

The simulations were conducted following the standard way using a Metropolis algorithm with the jackknife and binning methods~\cite{montecarlo-1,montecarlo-2}  in order to  evaluate the errors on the expectation values \eqref{expet}. The usual update algorithm, used to generate a new configuration, changes all the entries of the independent matrices:
\begin{equation}
\varphi_{ij} \longrightarrow  \varphi_{ij} + a_{ij}, \ \textrm{with}  \ i,j=1,\cdots N .
\end{equation}
where $N$ is the matrix size and $a_{ij}$ are random real numbers. The amplitude of variation of each matrix entry is tuned dynamically during the run time, requiring an acceptance rate\footnote{The acceptance rate is defined as the number of new configuration accepted over the total number of configurations tried } between $10\%$ and $30\%$. The new proposed configurations are judged by the metropolis algorithm, which requires to compute the difference between  the action for the new proposed configuration and the action for the configuration at the current time step $\delta S_{ij}(x) = S(\varphi_{ij}) \to \varphi_{ij} + x)$. The probability of accepting  the new configuration $\varphi_{ij} + x$ is calculated as  $\min(e^{-\delta S}, 1)$. In order to reduce the computation time, instead a global matrix change, we have changed just one matrix entry for each  Monte Carlo step. The coefficient to update is not chosen randomly but following an order used to vary all the coefficients after $N^2$ time steps. The interaction term proportional to $\varphi^4$ of the potential in the action \eqref{S-Ncom}  couples  a matrix entry $\varphi_{ij}$  to all the other elements on its line and column. This means that using the previous described update procedure, the computation of the variation of the action $\delta S$ (and consequently the Metropolis check) needs a number of operations which increases linearly with the matrix size\footnote{The computation of quadratic part of the action, which couples each element of the matrix with another one and the calculation of the observables using the formula \eqref{expet} are always sub dominant in terms of the number of operations.}. The drawback of this approach are mainly two: the complication of the code and the dependence of the correlation time $\tau$ in $N$. In fact two configurations will share at least one coefficient until $N^2$ Monte Carlo steps, thus this correlations introduce an increase of the correlation time increasing $N$ and presumably $\tau$ is proportional to $N^2$. However, using this optimization and the implementation of parallel computing we were able to execute our simulation  for $N=20$ with  a good precision and in reasonable time despite our limited computation resource.
The initial conditions of the the Markov chain are chosen randomly in the configuration space although we used two general types; hot initial conditions, which are configurations far from the minimum and cold start conditions, which correspond to configurations close to the minimum. As usual we have excluded all the time step  before the thermalization process take place.

\subsection{Order parameters}

The previous introduced quantities are not sufficient if we want to
measure the various contributions of different modes of the fields
to the configuration $\varphi$. We need some control parameters,
usually called order parameters. Recalling the fuzzyfication
procedure we associate to a function on the disc a matrix given by
coefficients $f_{nm}$ in the density matrix base expansion of~\eqn{densitymatrixbasis}: 
\be
\varphi(z,\bar{z})=e^{-\frac{z \bar{z}}{\theta}}\sum_{m,n=0}^\infty \varphi_{nm}\frac{z^n \bar{z}^m}{\sqrt{m!n!\theta^{n+m}}} =e^{-\frac{\rho^2}{\theta}}\sum_{m,n=0}^\infty \varphi_{nm}\frac{\rho^{n+m}e^{i\vartheta(n-m)}}{\sqrt{m!n!\theta^{n+m}}},\label{dens-r}
\ee
Due to the linearity of the expansion and since the action is even, the coefficients $\varphi_{nm}$ have the property to average to zero. As a first idea, we can think about a quantity related to  the square norms of the fields, for example the sums $\sum_{nm} |\varphi_{nm}|^2$. This quantity is
called the full-power-of-the-field; it can be computed as the trace of the square:
\be
\varphi^2_a = \operatorname{Tr}(\varphi^2) \label{vara} 
\ee
The quantity $\langle\varphi_a\rangle$ alone is not a good order parameter because it does not distinguish contributions from the different modes, but we can use it as a reference to define the quantity:
\begin{align}
\varphi^2_0 &= \sum^N_{n=0} |\varphi_{nn}|^2 \;, \label{var0}
\end{align}
it is easy to see that the quantity~\eqref{var0} is connected with the purely radial contribution, thus can be used to analyze the pure radial contribution to the full-power-of-the-field.
We will see that $\langle\varphi^2_0\rangle \approx 0$ defines a disordered phase while $\langle\varphi^2_0\rangle > 0$ defines a kind of ordered regime. In order to describe the type of ordering it is necessary to study the contribution from higher modes to $\varphi^2_a$.

We can generalize the previous quantity and define parameters $\varphi_l$ in such a way that they form a decomposition of $\varphi^2_a$:
\be
\varphi^2_a=\varphi^2_0+\sum_{l>0} \varphi^2_l;.
\ee
Following this prescription, the other quantities for $l>0$ can be  defined as:
\begin{equation}
\varphi^2_l =\sum^{N-l}_{n=0}\left(|\varphi_{n,n+l}|^2+|\varphi_{n+l,n}|^2\right) \;.\label{varl}
\end{equation}
If the contribution is dominated by the radial symmetric parameter we expect to have $\langle\varphi^2_a\rangle \sim \langle\varphi^2_0\rangle$.
In the next simulations we will evaluate, apart from $l=0$, the quantity with $l = 1$ as representatives of those contributions where
the rotational symmetry is broken.  According to  \eqref{varl} we have
\be
\varphi^2_1 =\sum^{N-1}_{n=0}\left(|\varphi_{n,n+1}|^2+|\varphi_{n+1,n}|^2\right) \, \;.
\ee
Using higher $l$ in \eqref{varl} we could analyses the contributions of the remaining modes, but it turns out that the measurements of the first two modes are enough to characterize the behavior of the system.

\subsection{Phase transitions}

Before we start let us remind of some aspects of the 2-dimensional $\varphi^4$
model on a fuzzy sphere. Many simulations have focused on scalar fields on fuzzy spheres  and they lead show the presence of new phases among the classical ones, characterized as nonuniform (or matrix or striped) ordered phase \cite{Medina,Fuzzy-numer-1,panero-1,Fuzzy-numer-2,Fuzzy-numer-3,Fuzzy-numer-4}. 

Numerical studies show three different phases: a disordered phase and an ordered regime, which splits into phases of uniform and nonuniform order. In the nonuniform phases, which  is  not
present in the commutative planar $ \varphi^4$ theory, the rotational symmetry is spontaneously broken. The three
transition curves intersect at the so called triple point. Analyzing the  limit $N \to \infty$ limit (which for the fuzzy sphere leads to the noncommutative plane) is been found  that the three curves and the triple point collapse using the same scaling function of $N$. Therefore, the triple point, all three phases and in particular the new nonuniform  phase,  survive in the limit. This means  that the naive scalar field action  can not be used as an approximation of the scalar field theory on the commutative 2-sphere. Using a perturbative approach \cite{Fuzzy-Action-mod} it is possible to prove  that in the limit of the commutative sphere  the two point function is non affected by the UV/IR mixing but again the  fuzzyfication does not commute with the commutative limit. Taking a  different limit to the noncommutative 2-plane the  UV/IR mixing reappears. So we can see the appearance of the nonuniform phase as a finite version of the UV/IR mixing.

These nonuniform phases are closely connected with spontaneous symmetry breaking \cite{Fuzzy-numer-5,Fuzzy-numer-6},  this appearance of the new phase seems in contradiction with the Coleman-Mermin-Wagner theorem which states that there can be no spontaneous symmetry breaking of continuous symmetry on 2-dimensional commutative spaces. However the theorem is strongly based on the locality of interactions therefore is no direct generalization of the  theorem for the  noncommutative case, where the theory is nonlocal. In the work of Gubser and Sondhi \cite{striped}, where those new  phases were originally conjectured, it is argued
 that this translation non-invariant phase is possible only in dimensions $d \geq 3$. Nevertheless, subsequent
numerical simulation,  in particular of Ambjorn and Catterall~\cite{striped-1}, show the existence of such a phase even in $d = 2$.

\begin{figure}[htb]
\vspace{-10pt}
\begin{center}
\includegraphics[scale=.60]{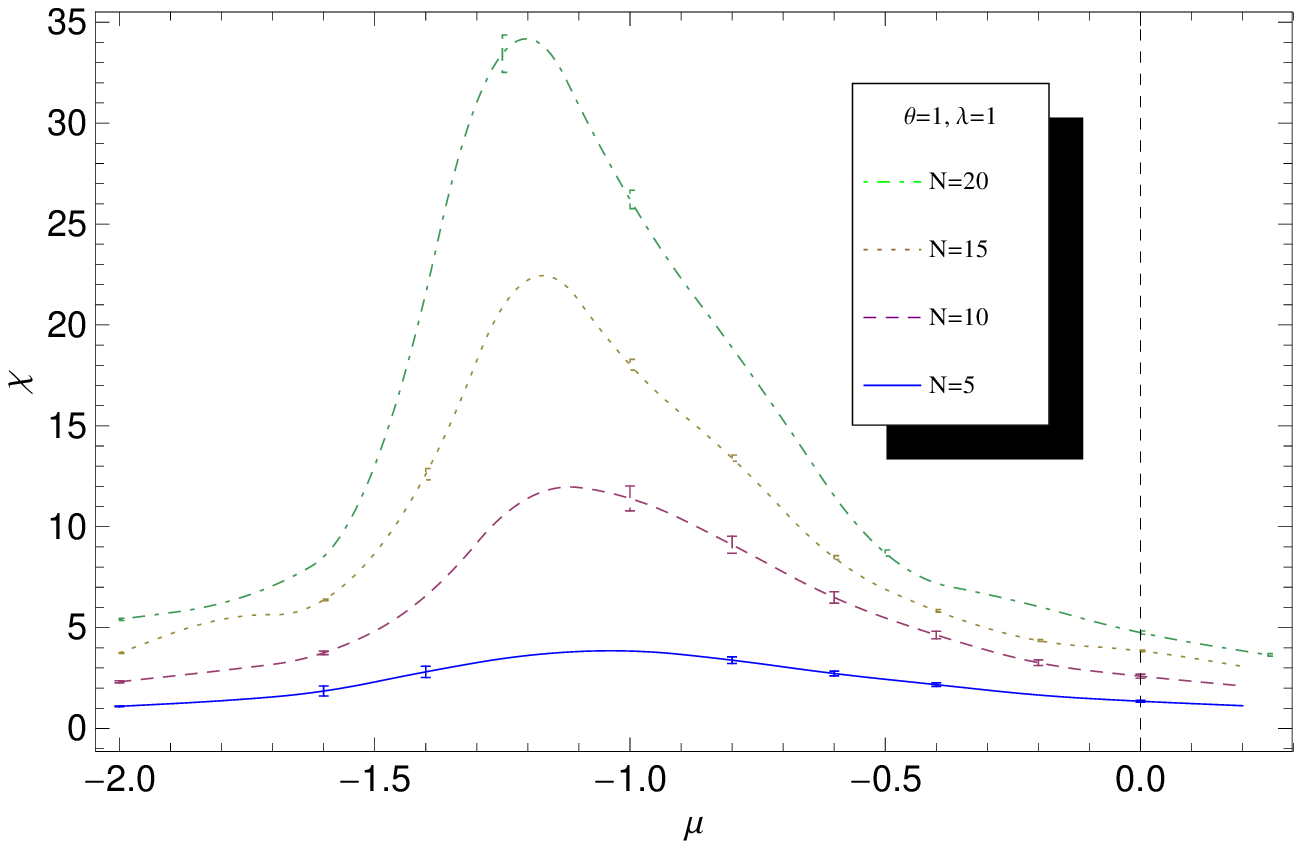}
\includegraphics[scale=.60]{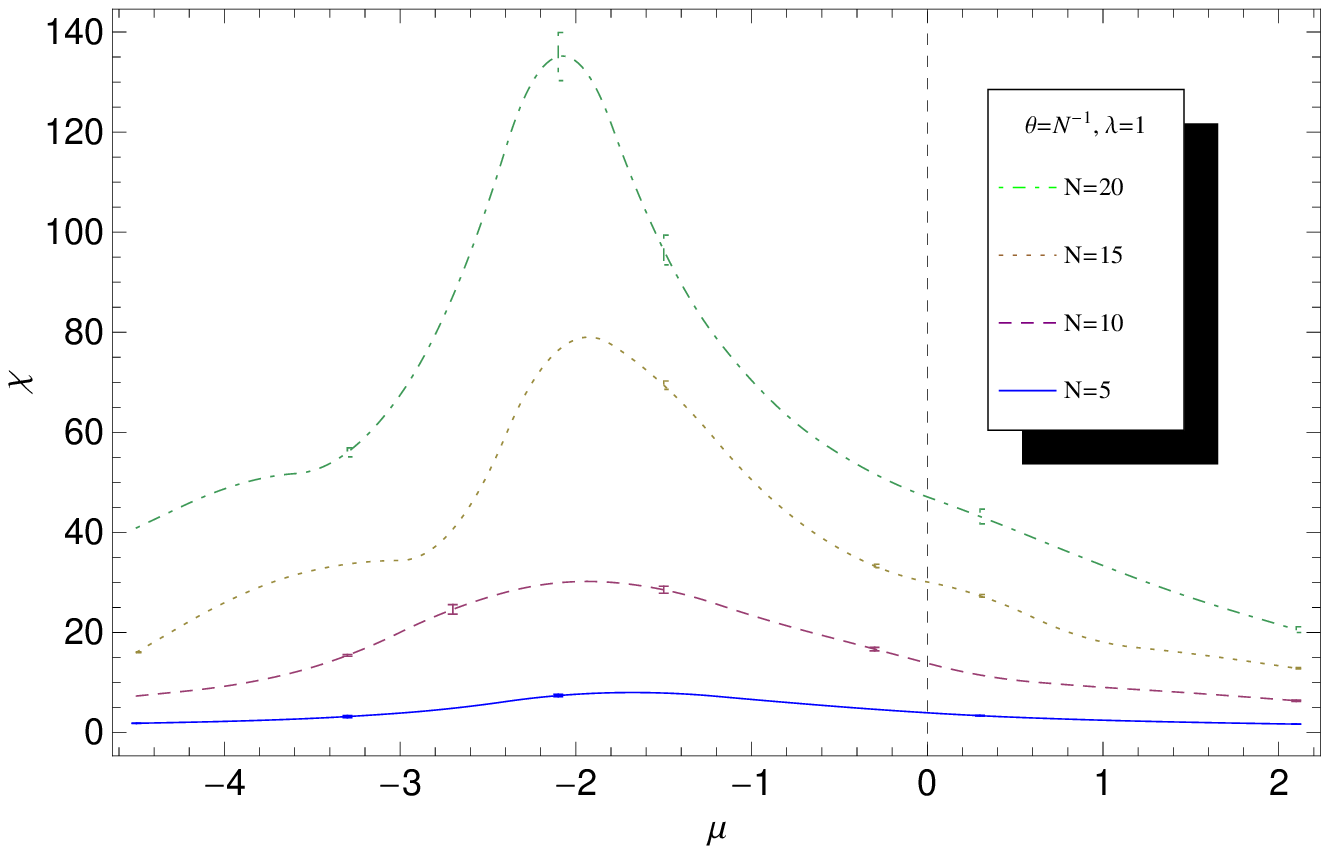}
\includegraphics[scale=.60]{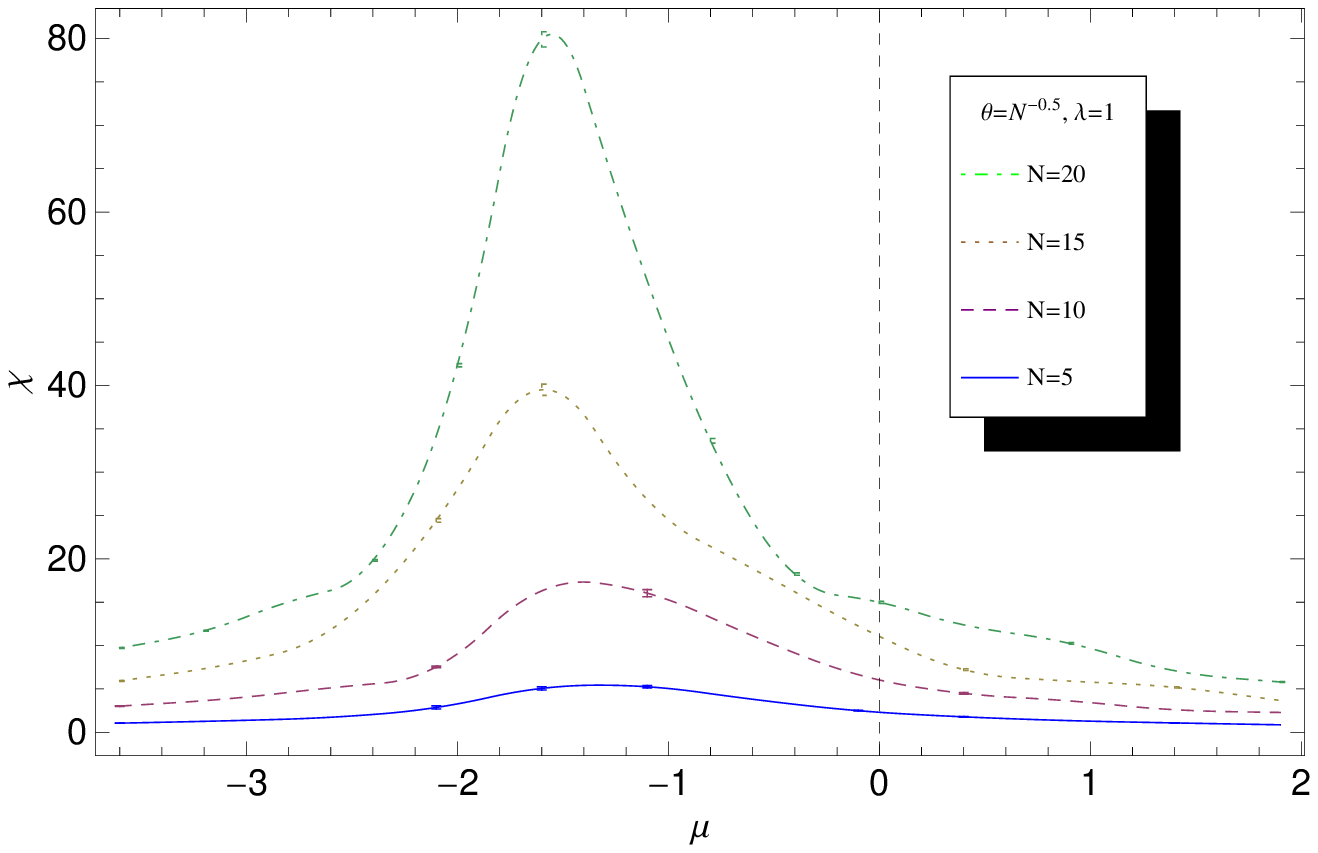}
\end{center}
\vspace{-30pt}
\caption{{\sl\footnotesize  Plots of the susceptibility in the different limits \CD, \NCP, and \CP ($\theta=1,N^{-1},N^{-05}$), for $\lambda=1$ and $N=5,10,15,20$.}
\vspace{-10pt}
\normalsize}\label{Figure1}\end{figure}

Now we are ready to discuss the result of the Monte Carlo simulations and  the various phases which arise in our model. In the following, for the first phase transitions, we use parameters $\lambda$ or $\mu$ of \eqref{par-1}. Fig.~\ref{Figure1}  shows the susceptibility \eqref{sus} plots for a fixed values of $\lambda=1$, varying the matrix size and for all the limits described. The location of a maximum in the susceptibility is used to identifies the critical value related to the phase transition for the continuous model. In Fig.~\ref{Figure1}  the maximum of the three susceptibility in the $N=20$ case can be easily extrapolated for $\mu_c$ negative equals to $-1.608\pm 0.006;-2.30 \pm 0.01 ;-1.56 \pm 0.01$ respectively for the noncommutative plane (\NCP), commutative disc (\CD), commutative plane (\CP) limit. In general we find the critical parameter $\mu_c$ for negative values and for $-\mu \lessapprox \lambda$. 

\begin{figure}[htb]
\vspace{-10pt}
\begin{center}
\includegraphics[scale=0.60]{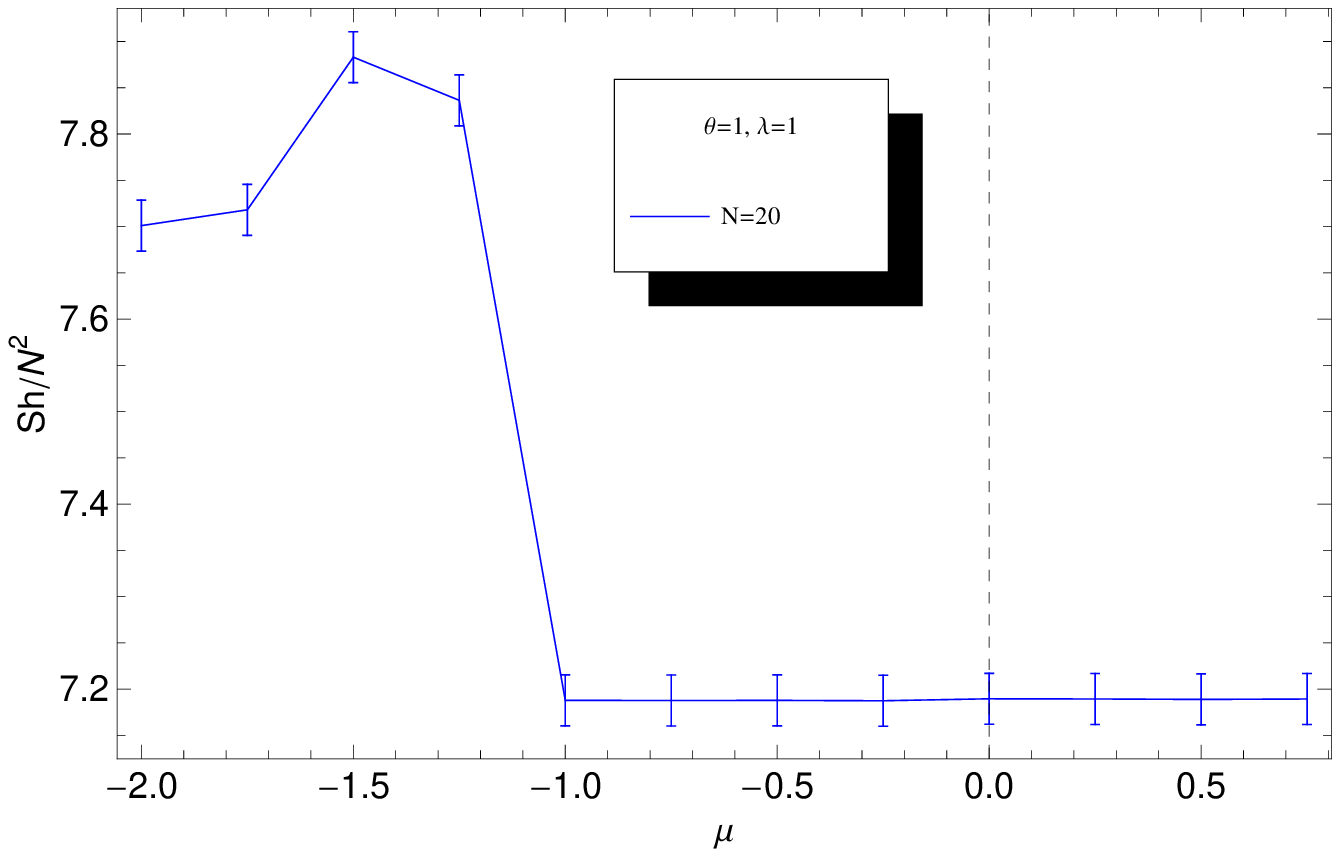}
\includegraphics[scale=0.60]{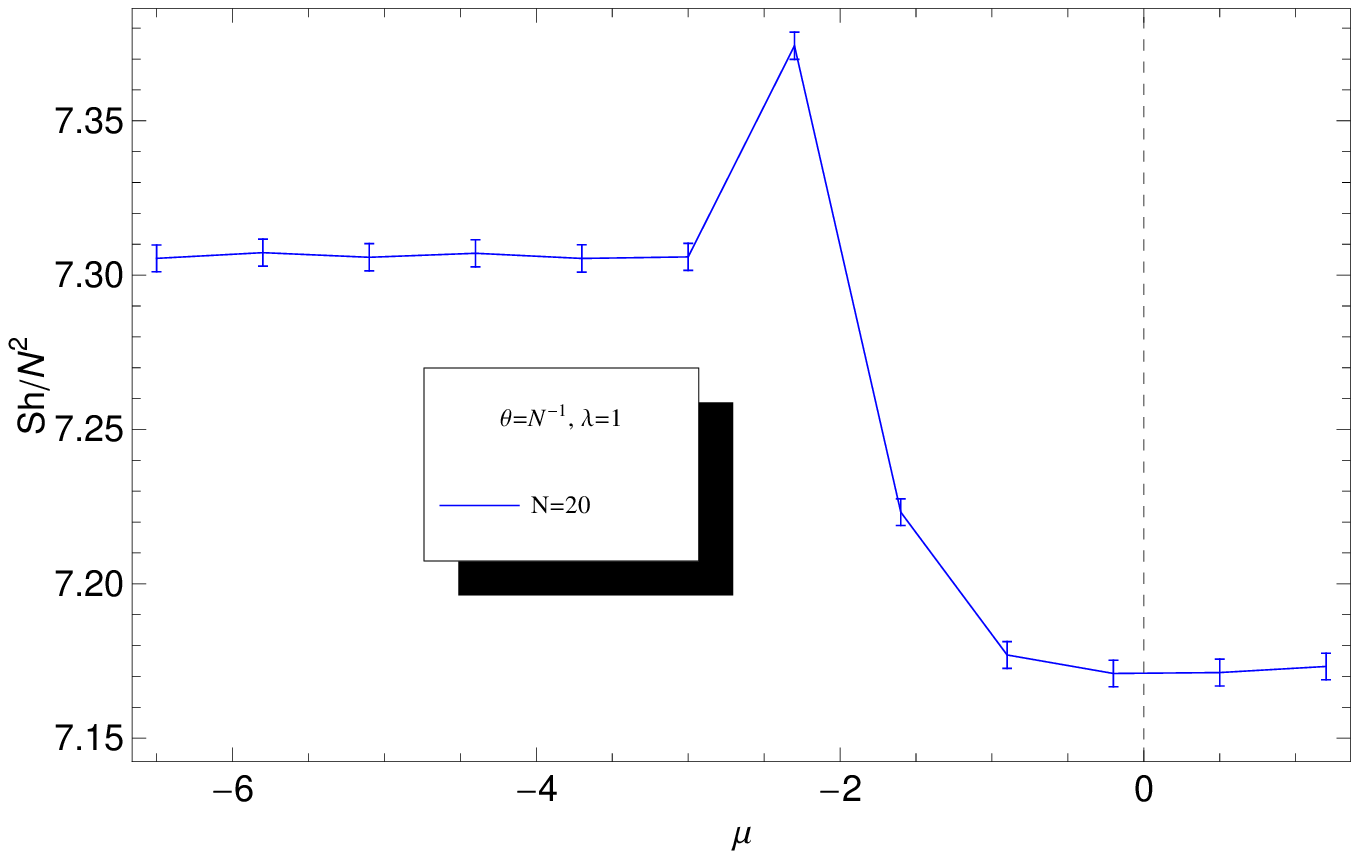}
\includegraphics[scale=0.60]{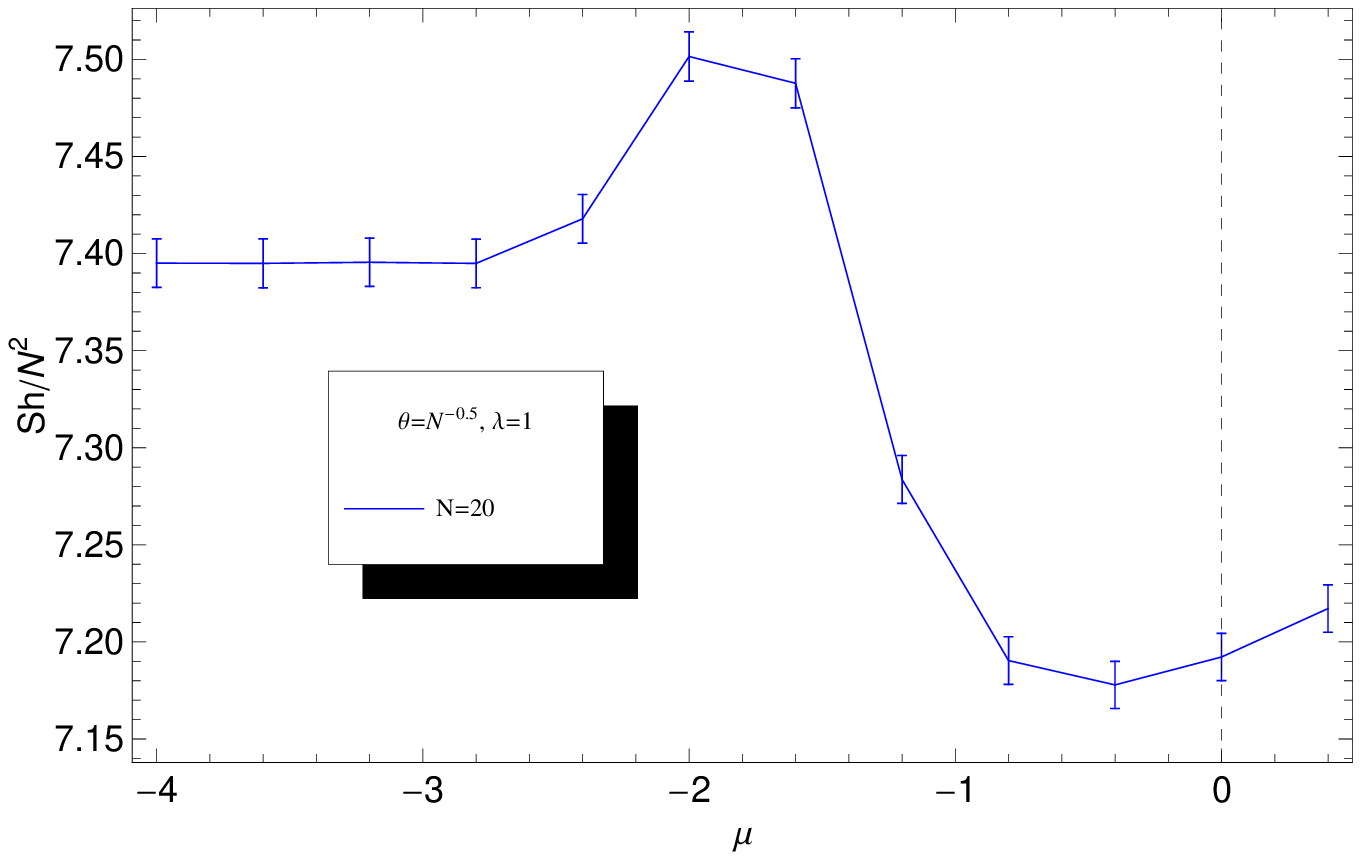}
\end{center}
\vspace{-30pt}
\caption{\sl \footnotesize  Plots of the specific heat density in the different limits \CD, \NCP, and \CP, for $\lambda=1$ and $N=20$
\vspace{-10pt}  \normalsize}\label{Figure2}\end{figure}

This critical behavior shows up in the specific heat density as well, in fact the specific heat can be used as an alternative in order to determine where the phase transition take place. Actually both quantities displays the same critical value but, comparing Fig.~\ref{Figure1}  and Fig.~\ref{Figure2}, we find a small difference in the critical value of $\mu_c$, likely due to a finite volume effect. In the following we adopt the the susceptibility criterion in order to find the transitions curves because it is easier to estimate the maximum of susceptibility respect to estimate the critical parameter through the specific heat step. Beside using the same statistic the errors are smaller for the susceptibility than in the case of the specific heat since in general more statistics is required for the specific heat to have the same precision. From the slopes of the specific heats and of the the susceptibility we can infer information about the order of the phase transition. In fact this phase transitions turn out to be  of the second order for all the three cases.

\begin{figure}[htb]
\vspace{-10pt}
\begin{center}
\includegraphics[scale=0.60]{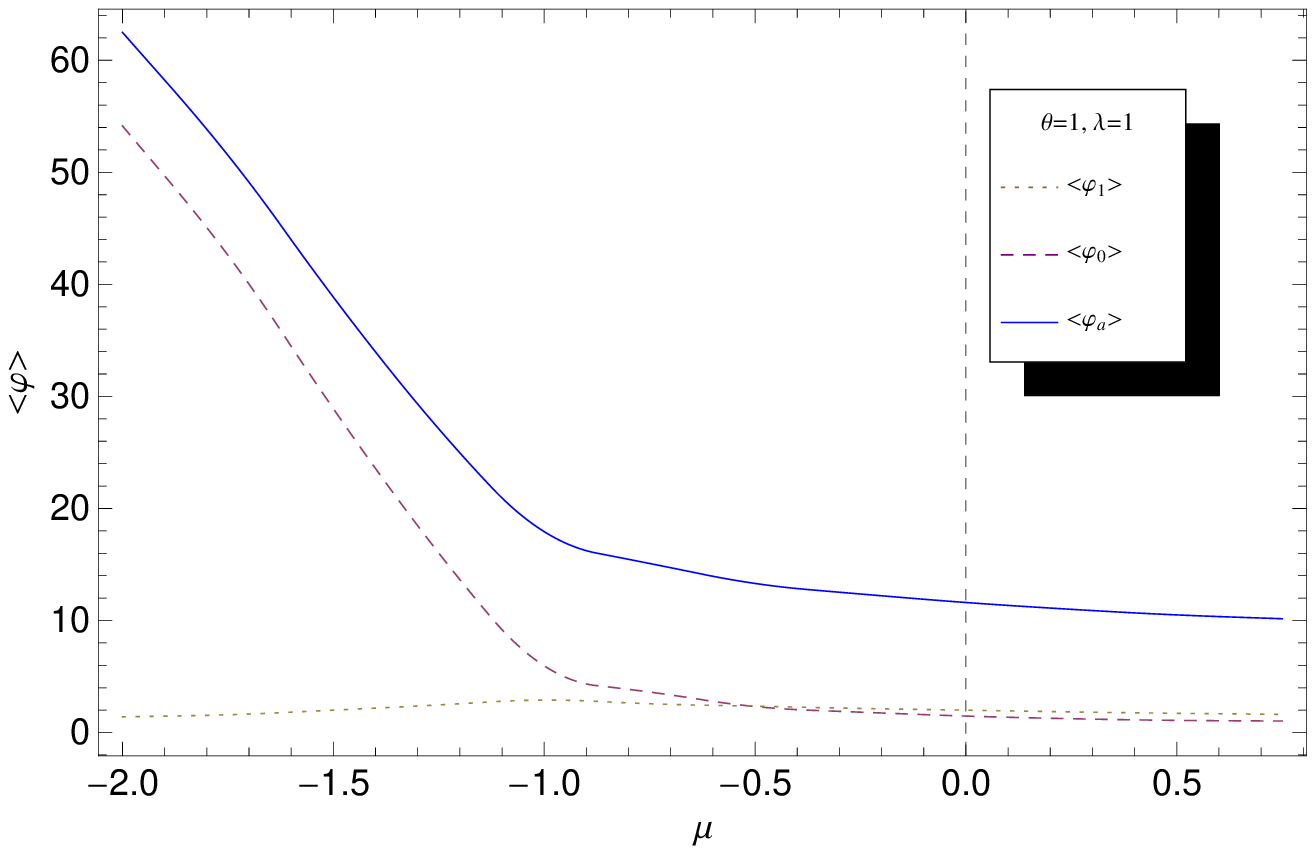}
\includegraphics[scale=0.60]{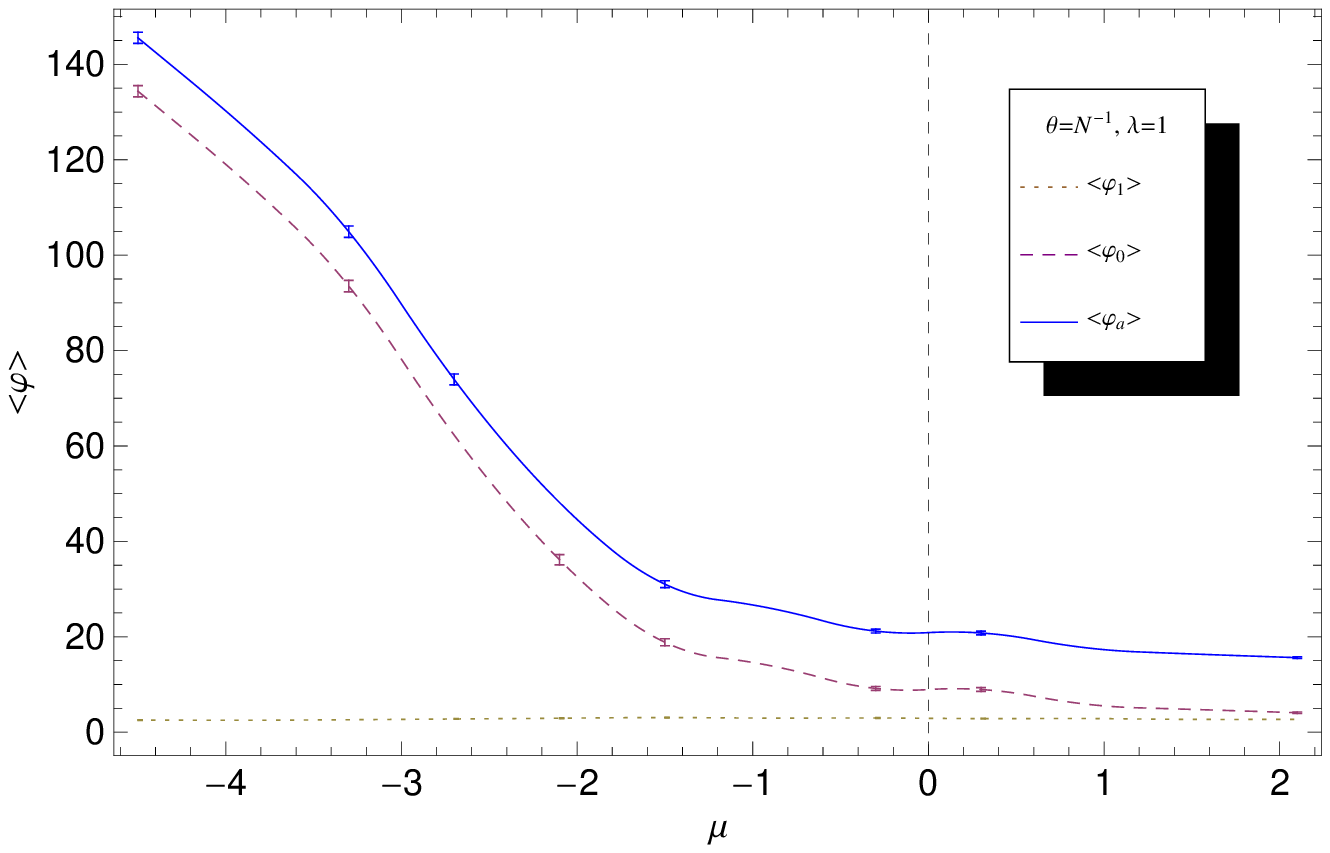}
\includegraphics[scale=0.60]{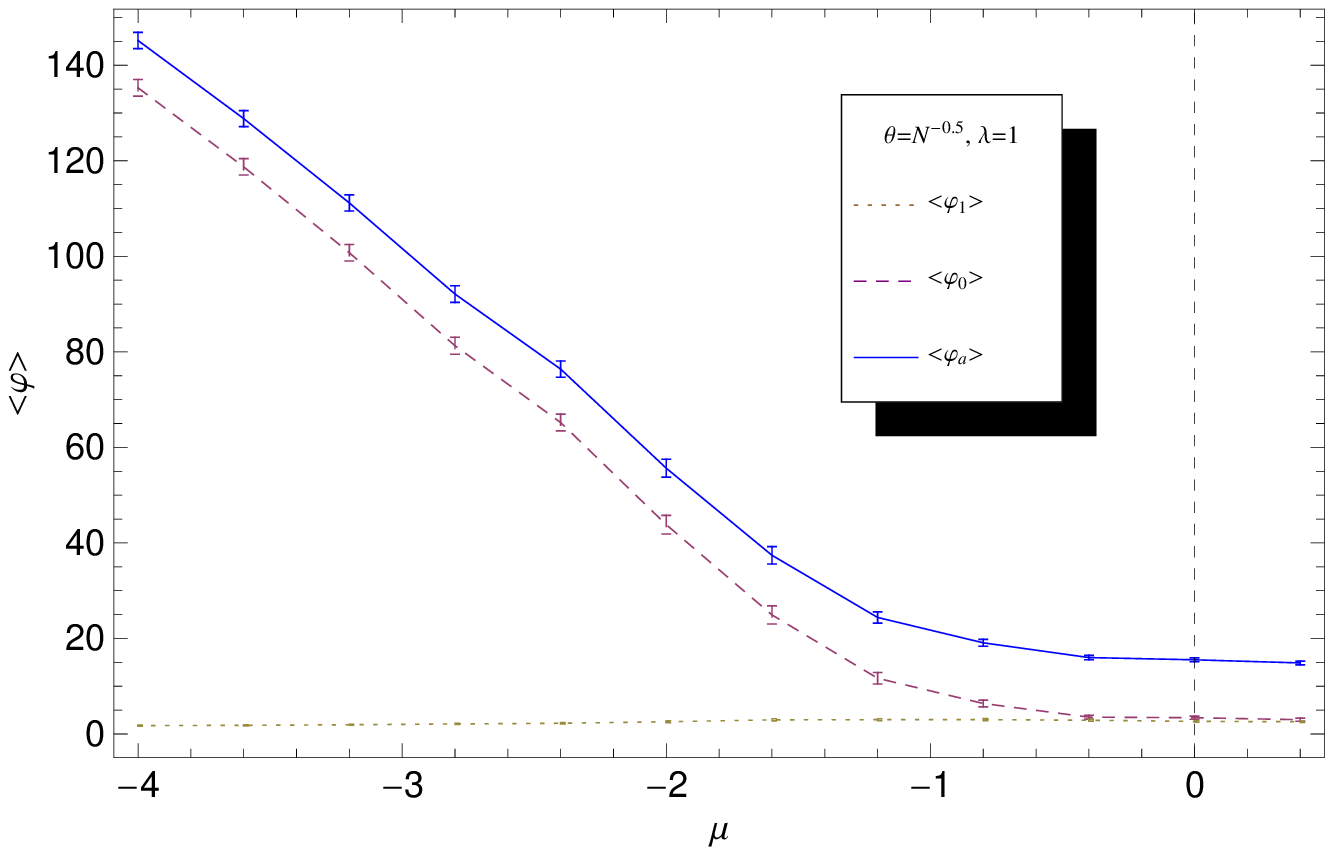}
\end{center}
\vspace{-30pt}
\caption{\sl \footnotesize Plot of the expectation values of  order parameters  in the different limits for for $\lambda=1$ and $N=20$ 
\vspace{-10pt}
\normalsize}\label{Figure3}\end{figure}

In order to gain some informations on the nature of the phases involved in the transitions we look at the order parameters defined in the previous section.  In Fig.~\ref{Figure3} are plotted the expectation values of the order parameters. Looking for values greater than the critical point $\mu > \mu_c$  we find  the $\langle\varphi^2_0\rangle $ and $\langle\varphi^2_1\rangle $  much smaller than the full power of the field $\langle\varphi^2_a\rangle$ with a weak dependence on $\mu$.

The behavior suggest that all the modes contributes effectively to the full power of the field as random fluctuations. This type of phase is usually named disordered phase due to the contribution of the complete composition of the field  by random fluctuations around the null constant average field.
For smaller values of $\mu $ with respect to respect the critical values   $\mu < \mu_c$
we find a  phase characterized by $\langle\varphi^2_a\rangle \gg 0$, the pure radial contribution $\langle\varphi^2_0\rangle $ is dominant while  $\langle\varphi^2_1\rangle \approx 0$ thus only one mode (the zero one) is mainly present in the field decompositions governing its behavior. In contrast with the previously described phase, where the  full powers of the fields are basically a sums of random fluctuations, this ones is named uniform phase. In this regime the our fields are essentially diagonal matrices and the kinetic action part of the action by the regions around its minima at $\varphi_\pm = \pm\sqrt{-2\mu/\lambda} $.
These results are completely consistent with what one would expect for the usual plane so we can state that all the three limits possess the disordered-uniform  phase transition observed for a $\varphi^4$ scalar field theory on the commutative plane.
 \begin{figure}[htb]
\vspace{-10pt}
\begin{center}
\includegraphics[scale=0.60]{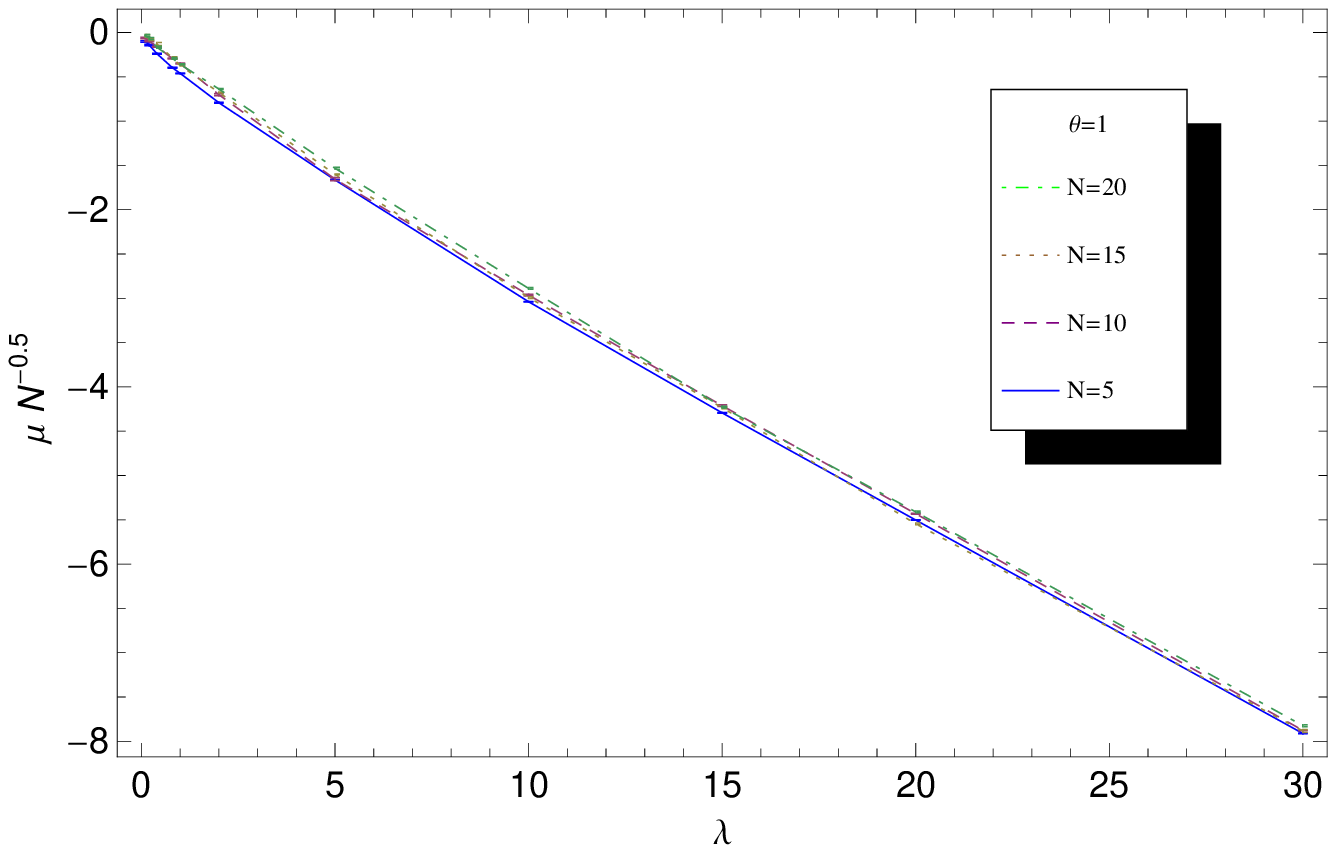}
\includegraphics[scale=0.60]{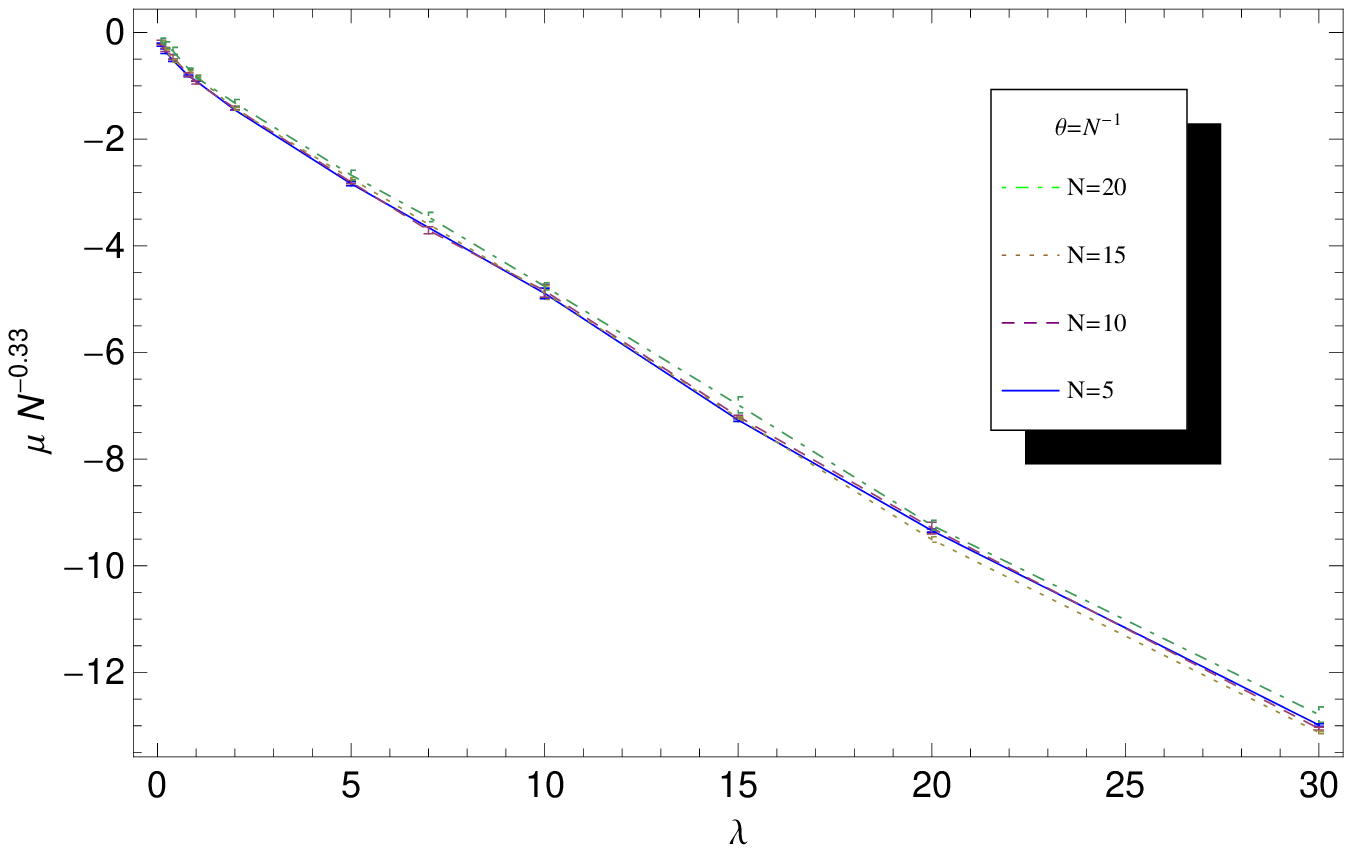}
\includegraphics[scale=0.60]{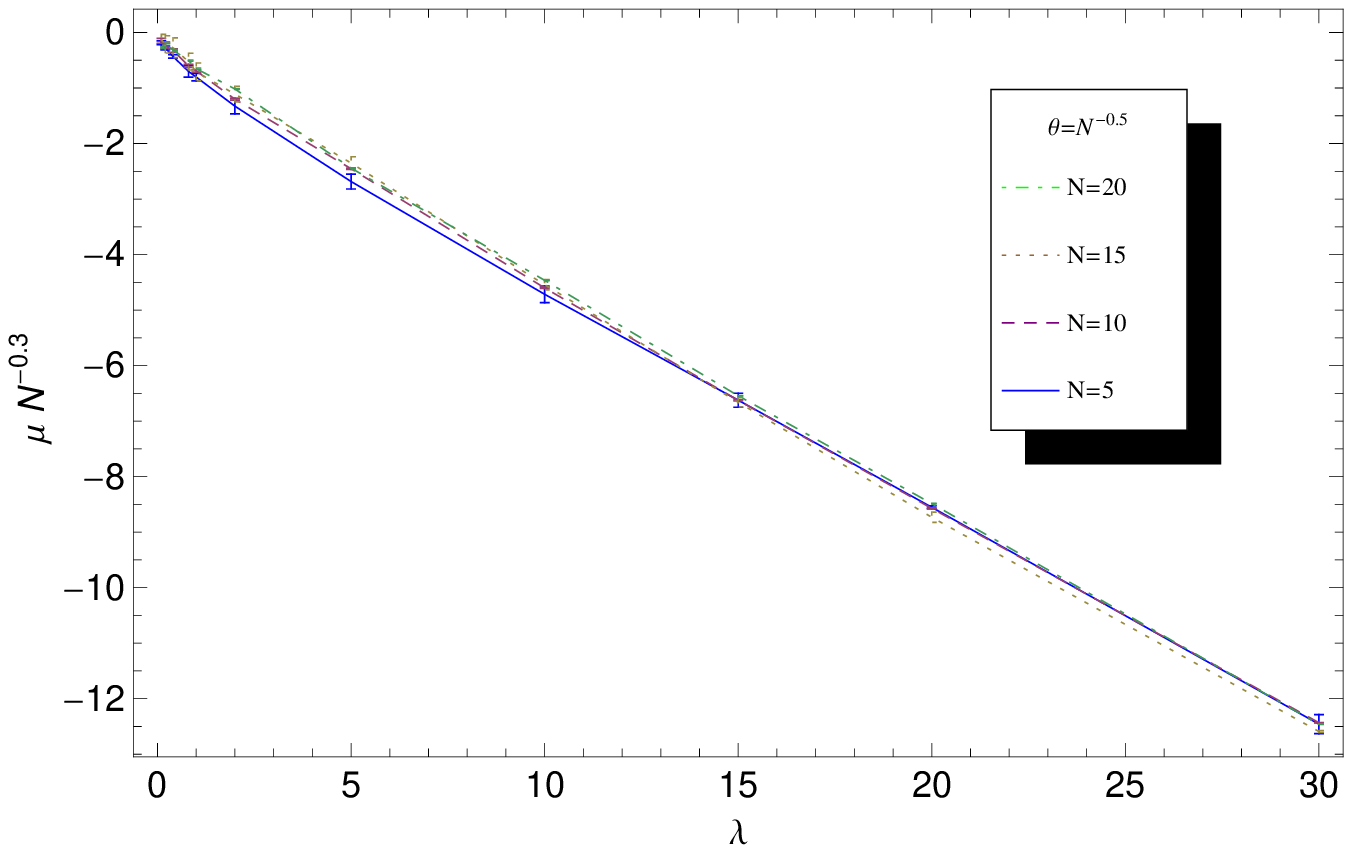}
\end{center}
\vspace{-30pt}
\caption{\sl\footnotesize Plot of scaled uniform-disordered transition curves  in the different limits  for  $N=5,10,15,20$
\vspace{-10pt}\normalsize}\label{Figure4}\end{figure}

Repeating the previous analysis for different values of $\lambda$ and finding the corresponding  critical values $\mu_c$ we can plot the transitions curves disorder-uniform; the  scaled plots are shown in Fig.~\ref{Figure4}. All the graphs possess  good scaling properties  even for small matrix sizes without strong finite dimensional effects. Each plot has a different scaling coefficient for the noncommutative plane limit the lines collapse for a scaling like the square root of the matrix size, for the commutative disc limit the best collapse take place for $N^{-0.33}$ and for the commutative plane for $N^{-0.3}$. Furthermore the phase transition line is well approximated by a linear function therefore  performing  a fit of the data  we obtains  the approximations:
\be
\mu^{\NCP}_c(N,\lambda)\approx -0.29 \lambda N^{-0.5}, \  \  \mu^{\CD}_c(N,\lambda)\approx -0.49 \lambda N^{-0.33}, \  \ \mu^{\CP}_c(N,\lambda)\approx -0.43 \lambda N^{-0.3}.
\ee
The previous linear regression has been done on the highest  matrix size data set $N = 20$, which is our best approximation to the large $N$ limits.
\begin{figure}[htb]
\vspace{-10pt}
\begin{center}
\includegraphics[scale=0.60]{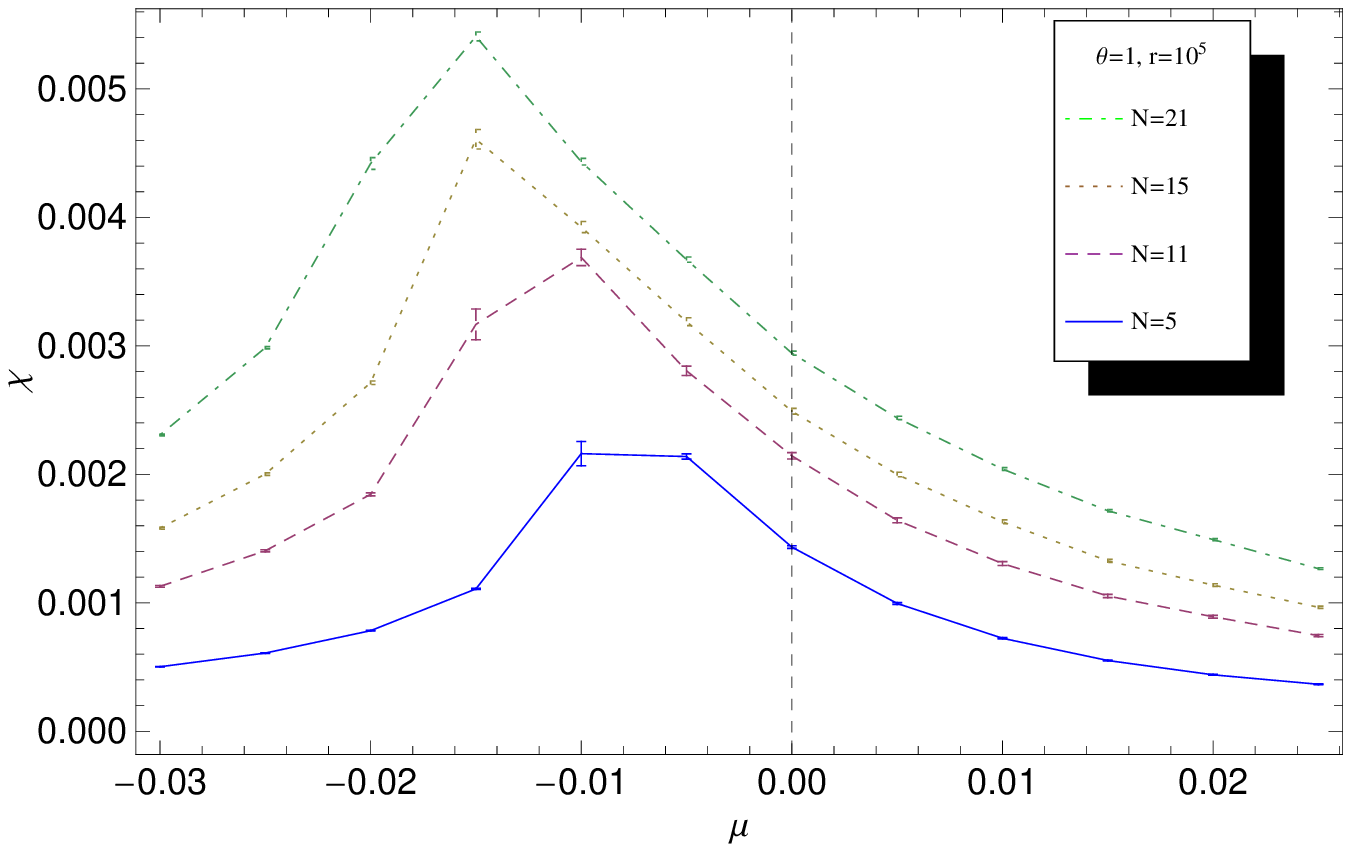}
\includegraphics[scale=0.60]{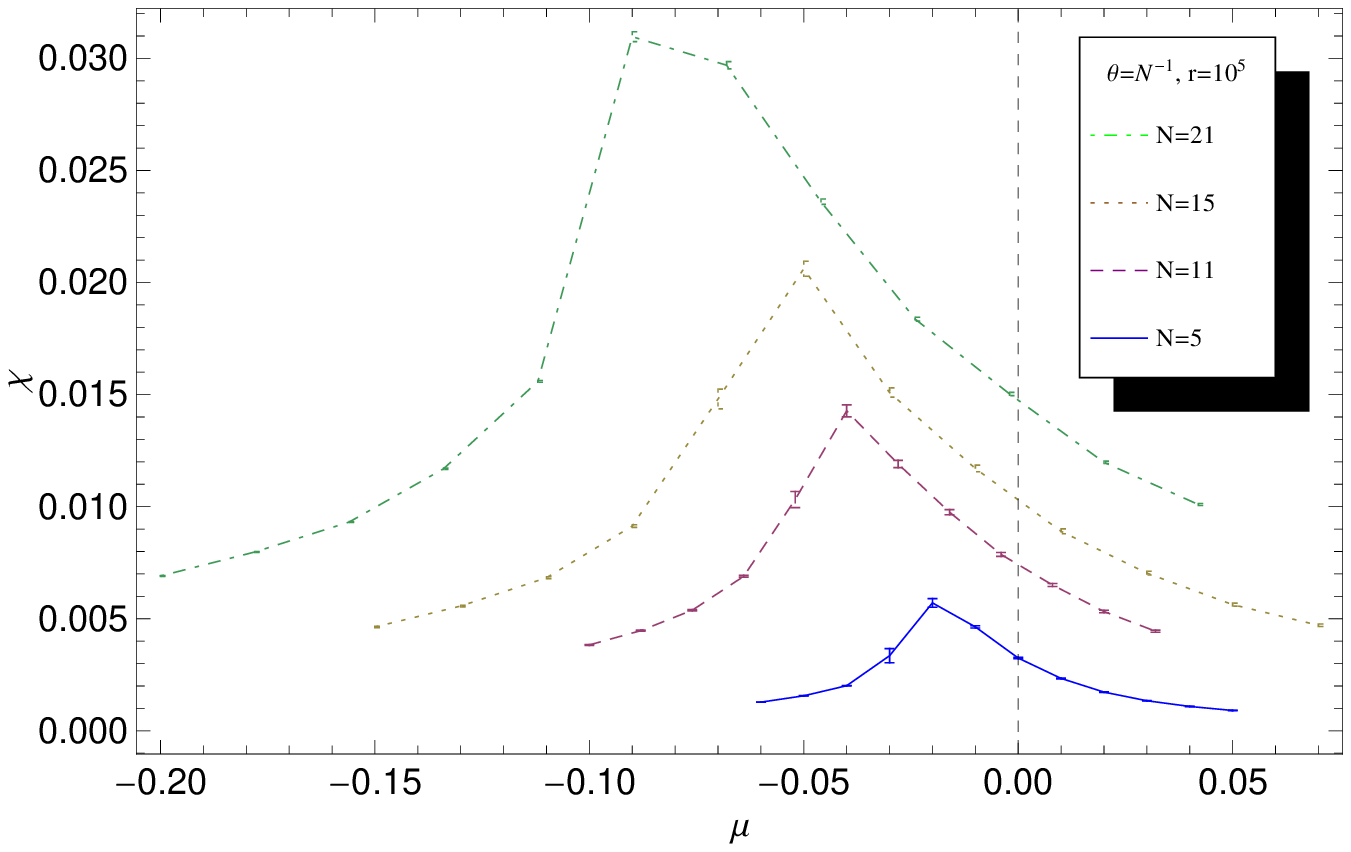}
\includegraphics[scale=0.60]{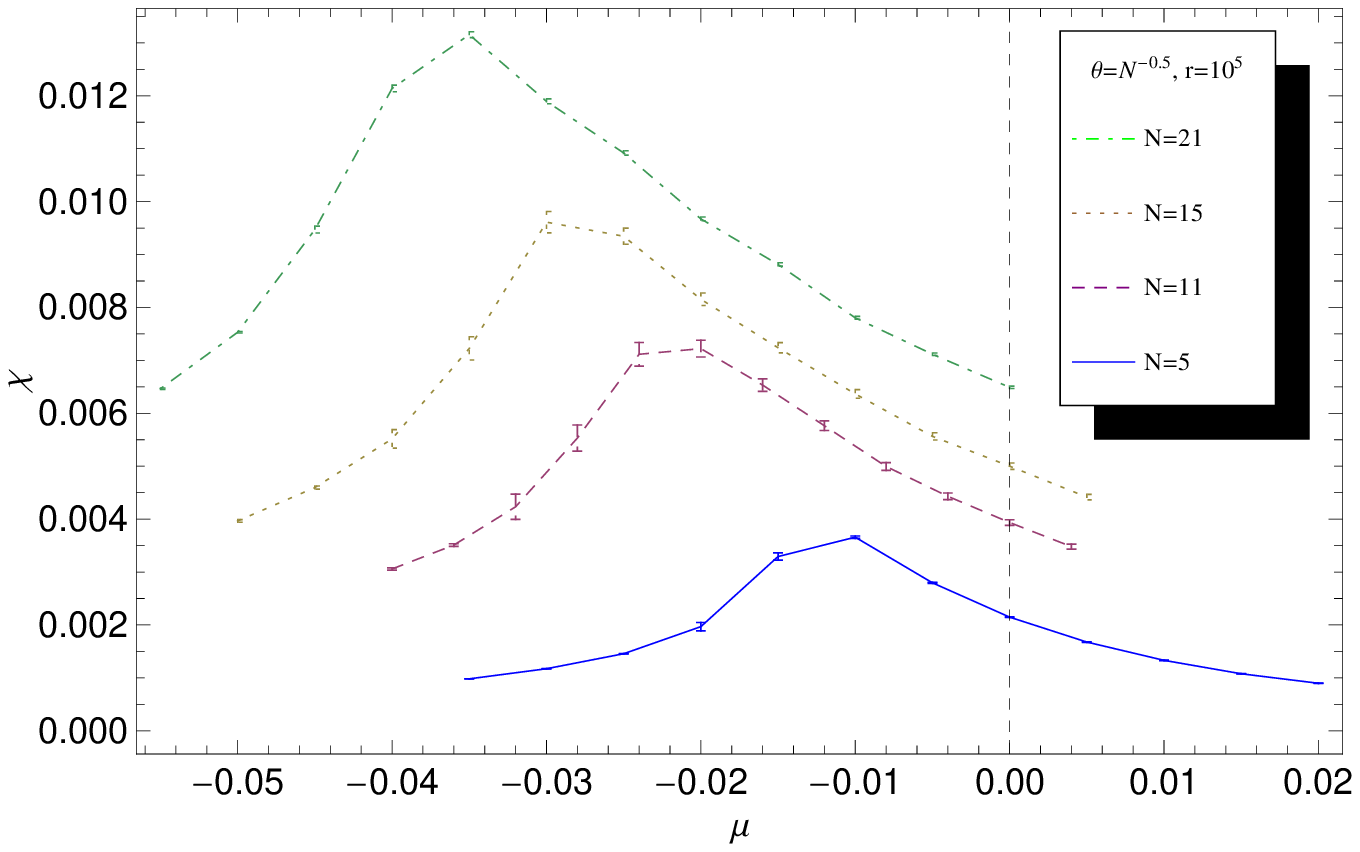}
\end{center}
\vspace{-30pt}
\caption{\sl\footnotesize Plots of the susceptibility in the different limits $\theta=1,N^{-1},N^{-0.5}$, for $r=10^5$ and $N=5,10,15,20$. 
\vspace{-10pt}\normalsize}\label{Figure5}\end{figure}
 We see that the two commutative case, \CD \ and \CP \ behave in a similar way among themselves, and are different from the noncommutative plane limit. 
Beside the uniform and disorder phases the simulations show the appearance of a new phase, in this case it is convenient to adopt the other choice of parameters \eqref{par-2}. We find this new phase transition for large $r$, Fig.~\ref{Figure5} show an  example of this phase transition for fixed  $r=10^4$ and varying $\mu$. Again the  critical points can be easy located using the maximum of the susceptibility. As we can see the locations of the peaks for all the plots are pushed to the left increasing $N$ in particular if we compare the plot concerning the non-commutative limit and the other two, we observe a bigger shift of the locations of the critical points to the left of the two commutative limit plots respect the noncommutative one. We will see this difference clearly in the transition curves.

\begin{figure}[htb]
\vspace{-10pt}
\begin{center}
\includegraphics[scale=0.60]{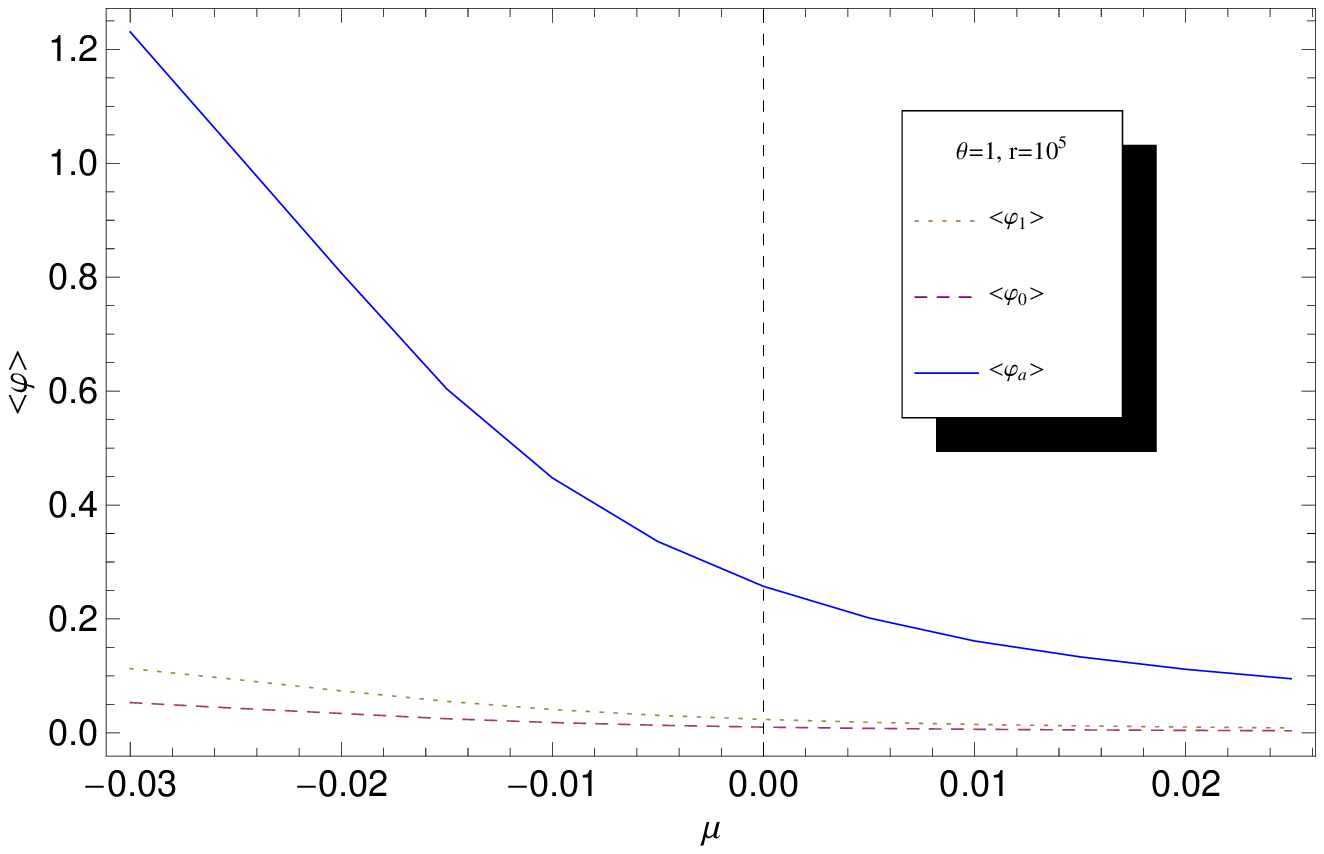}
\includegraphics[scale=0.60]{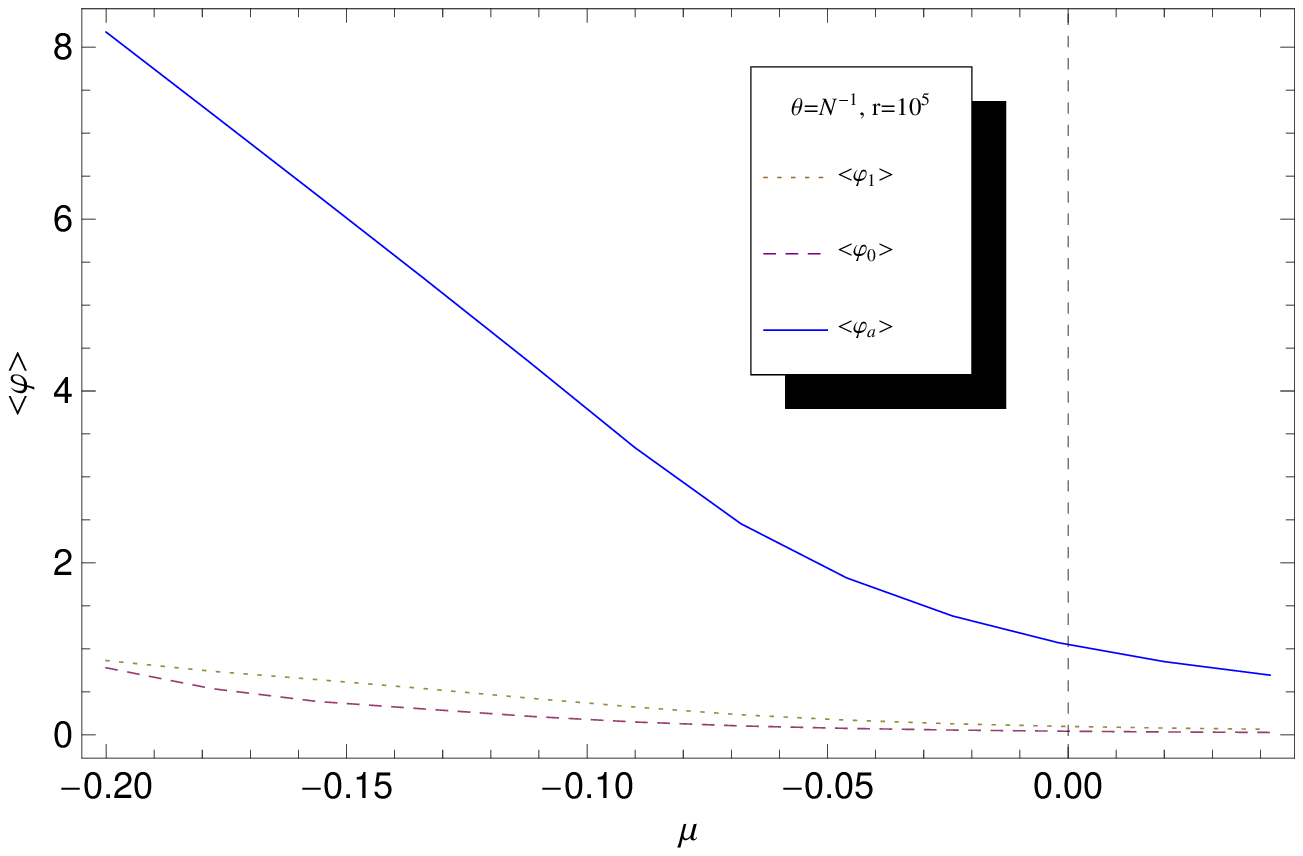}
\includegraphics[scale=0.60]{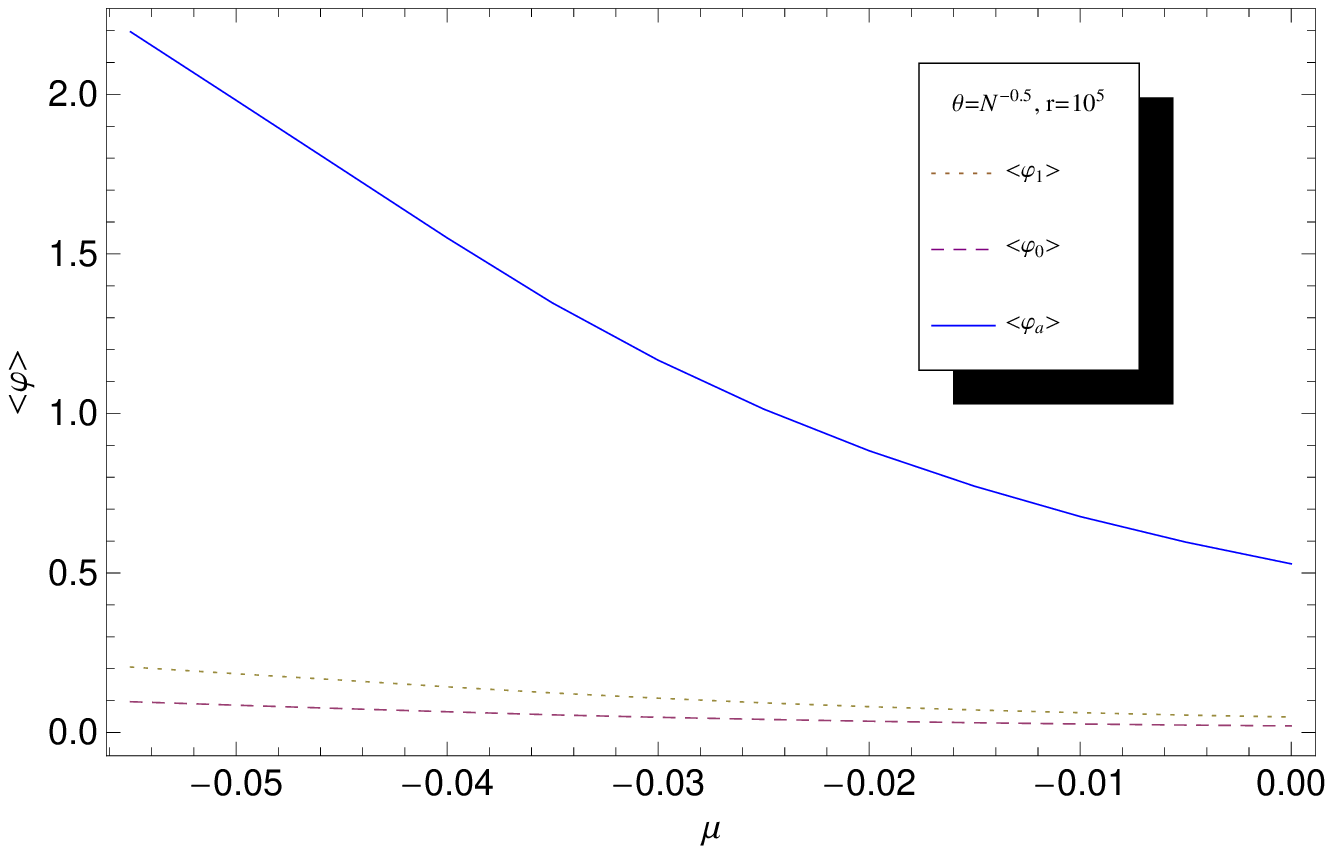}
\end{center}
\vspace{-30pt}
\caption{\sl\footnotesize  Plot of the expectation values of  order parameters  in the different limits for for $\lambda=1$ and $N=20$ \normalsize \vspace{-10pt}\normalsize}\label{Figure6}\end{figure}

Looking at Fig.~\ref{Figure6} the new phase appears characterized,  for $\mu<\mu_c$,  by $\langle\varphi^2_0\rangle  < \langle\varphi^2_1\rangle \ll \ \langle\varphi^2_a\rangle $ and $\langle\varphi^2_a\rangle  \gg 0 $, this is true for all the limits we have investigated, therefore we are dealing with the same phase. In particular we  notice that in the regions $\mu<\mu_c$  the full power of the fields becomes almost linear, after various fits on the data we obtain for all limits:
\be
\langle\varphi^2_a\rangle \approx  \frac{\pi}{N}\left( -2\pi \mu\right). \label{full-fit}
\ee
The factor $\frac{\pi}{N}$ is a normalization. We can generalize this result introducing another parametrization analogous to ~\eqref{par-2} in witch we reintroduce $\lambda$ but just as fixed parameter:
\be
S_V(\varphi) = \pi\theta r\operatorname{Tr}\left(\mu\varphi^2 + \frac{\lambda}{4}\varphi^4 \right).
\ee
In this case we have for the correspondent  full power of the field \eqref{full-fit} the fit:
\be
\langle\varphi^2_a\rangle \approx  \frac{\pi}{N}\left( -\frac{2\pi\mu}{\lambda}  \right). \label{full-fit-lam}
\ee
We recast our model in this way in order to show the analogies  with the fuzzy sphere \cite{Fuzzy-numer-1}, for which equation~\eqref{full-fit-lam} is very similar.
The behavior of the order parameters suggests that in the new phase the eigenvalues of the
field are in a minimum of the potential. Following \cite{Fuzzy-numer-1} the \eqref{full-fit-lam} hint that the neighborhood of
the minima of the potential  mainly contribute to the expectation values. The minima of the potential \cite{Fuzzy-numer-1} are $N$ disjoint orbits of the form
\be
O_n= \sqrt{\frac{-2\mu}{\lambda}} U^\dagger\left(1_n \oplus -1_{ N -n} \right)U  \ | \  U \in U(N)/[U(n) \times U(N - n)],
\ee
where $n \leq \frac{N}{2}$ and $1_n$ are the unit matrices. In the case of a negligible kinetic term,  the orbit with the largest phase space volume will dominated  the field composition. In the previous case of the uniform phase, the minimum of the potential is clearly found minimizing the kinetic term as well,  but in the new phase there is not a direct hint that the kinetic term might be negligible. To
test this assertion we have studied the partial contributions of the kinetic part $D$ and the potential part $V$ to the total energy.
\begin{figure}[htb]
\vspace{-10pt}
\begin{center}
\includegraphics[scale=0.60]{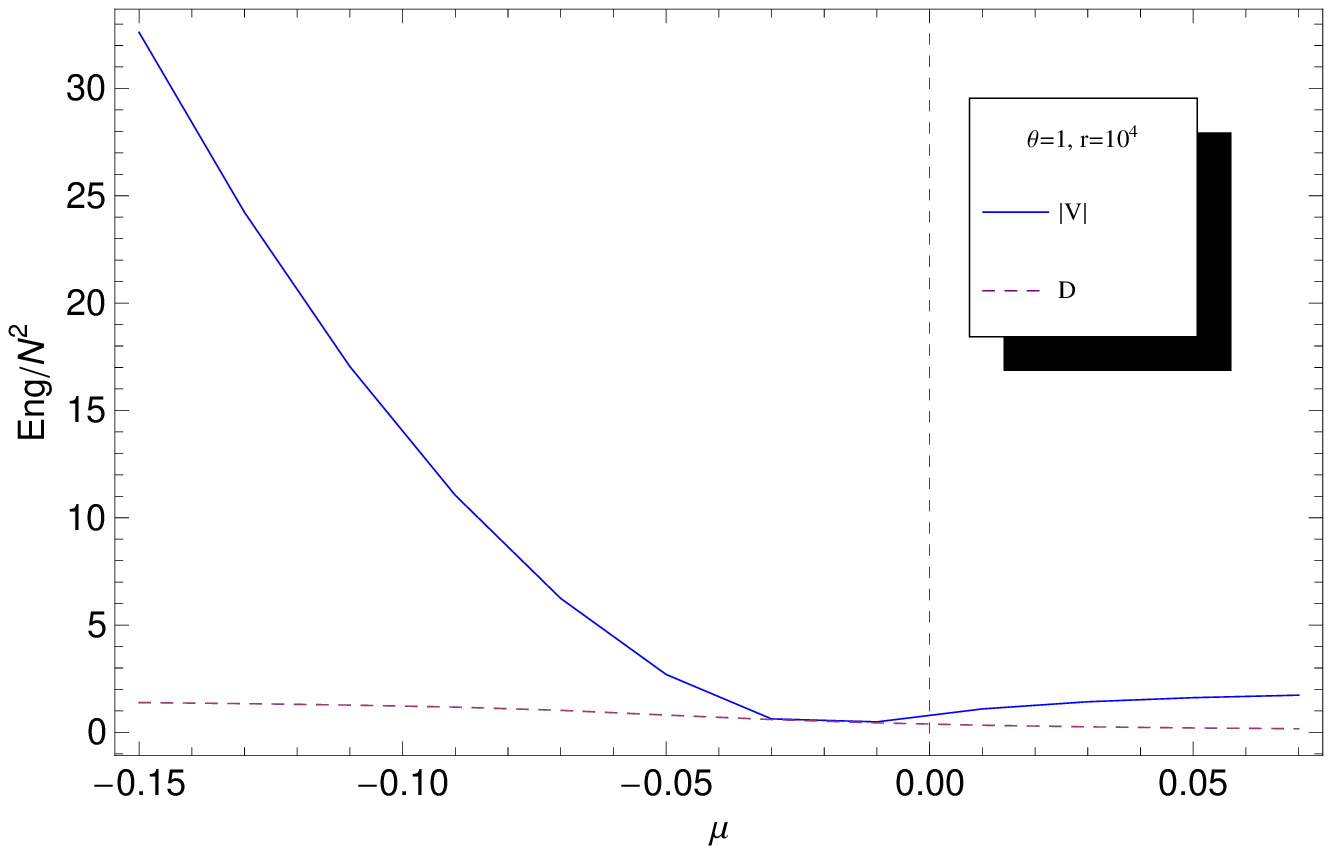}
\includegraphics[scale=0.60]{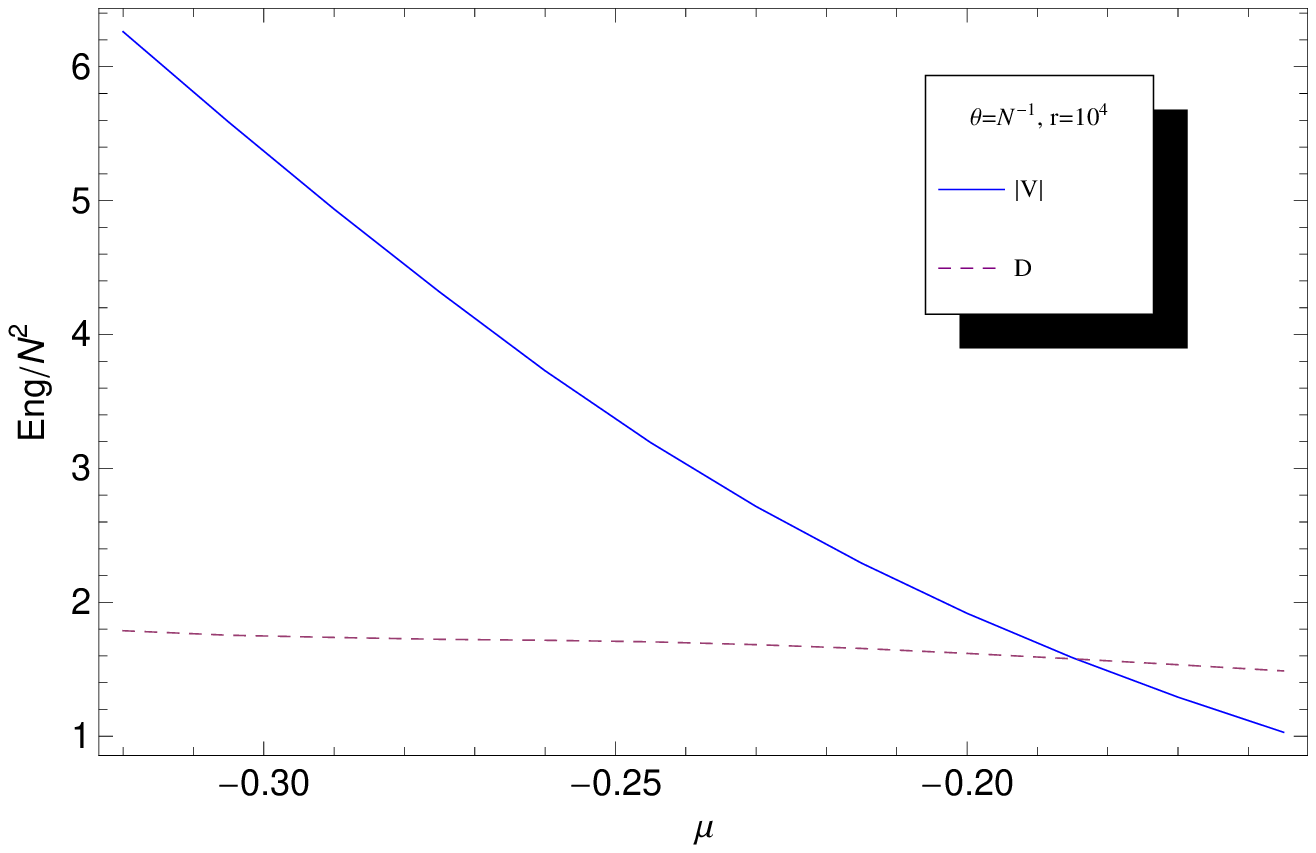}
\includegraphics[scale=0.60]{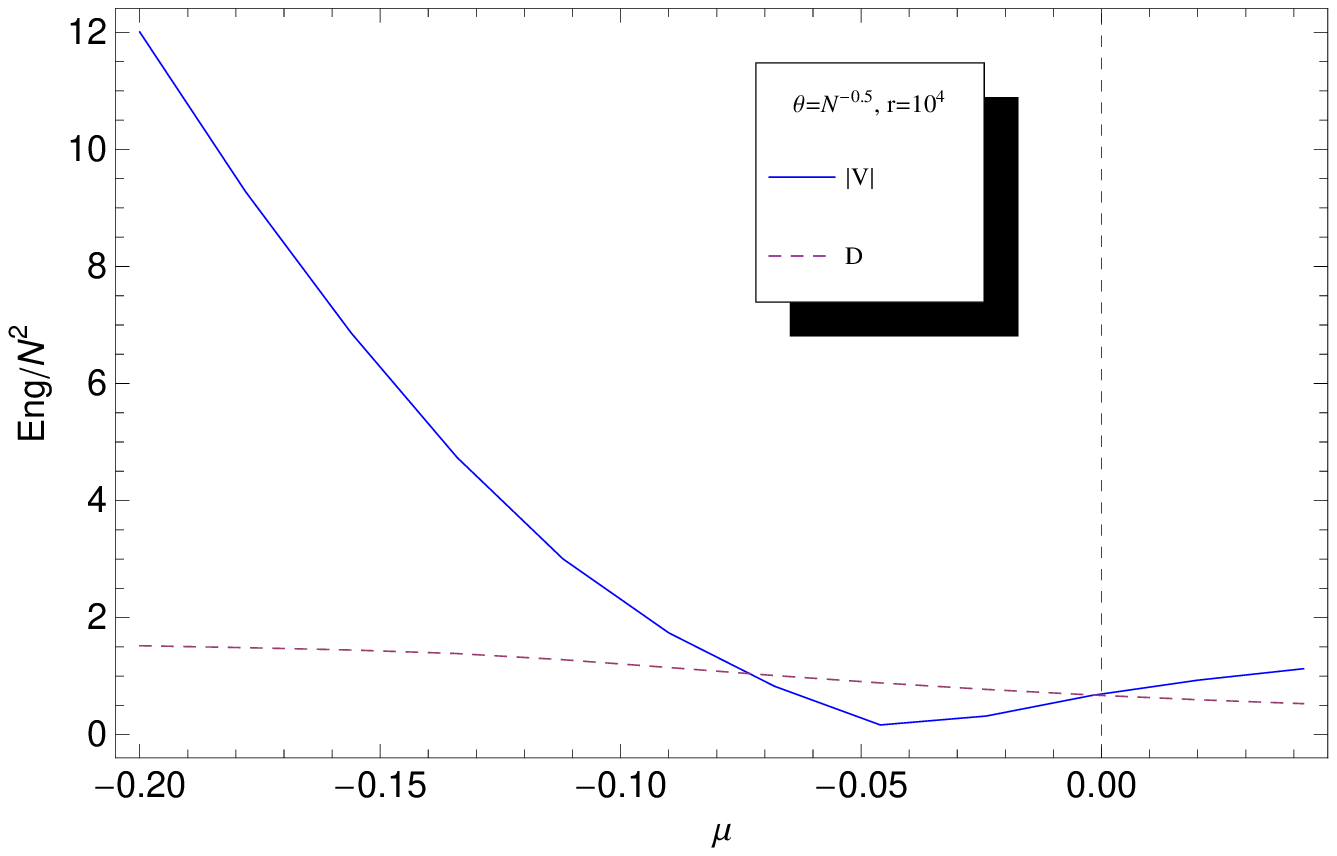}
\end{center}
\vspace{-30pt}
\caption{\sl\footnotesize  Plot of the  partial contributions $V$ and $D$ of  the internal energy $E$ in the different limits for for $r=10^4$ and $N=21$. \normalsize 
\vspace{-10pt}\normalsize}\label{Figure7}\end{figure}
As Fig.~\ref{Figure7} shows, our results confirm that for $\mu<\mu_c$ the transition the average kinetic contribution is much smaller than the modulus of the potential. However, this behavior is more marked for more  negative $\mu$-values  in the uniform order regime as well,  a fact that does not allow to interpret the rising of the this new phase caused only from the potential dominance. This new phase was already found for the same scalar field theory on the fuzzy sphere and is usually referred as non-uniform phase or matrix phase \cite{Medina,Fuzzy-numer-1,panero-1}.

\begin{figure}[htb]
\vspace{-10pt}
\begin{center}
\includegraphics[scale=0.60]{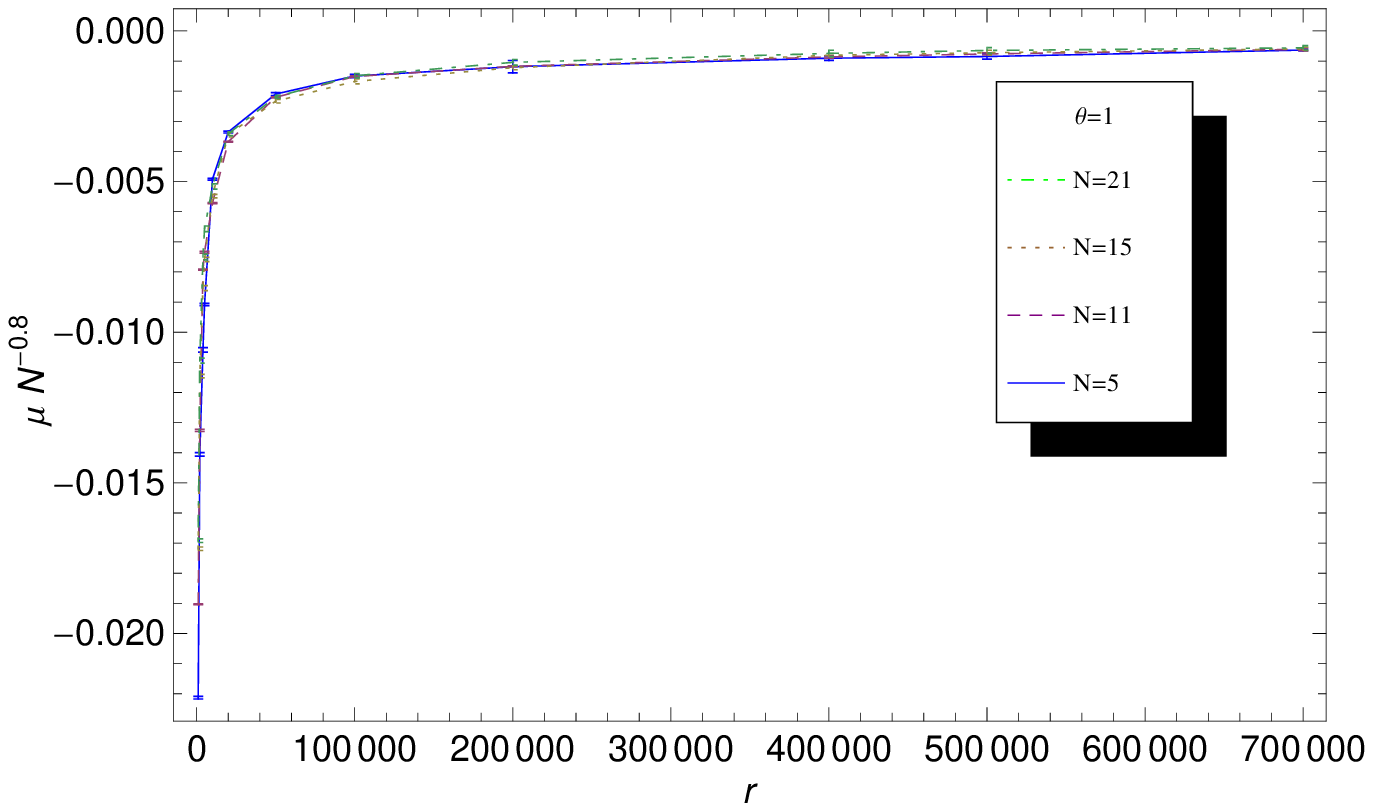}
\includegraphics[scale=0.60]{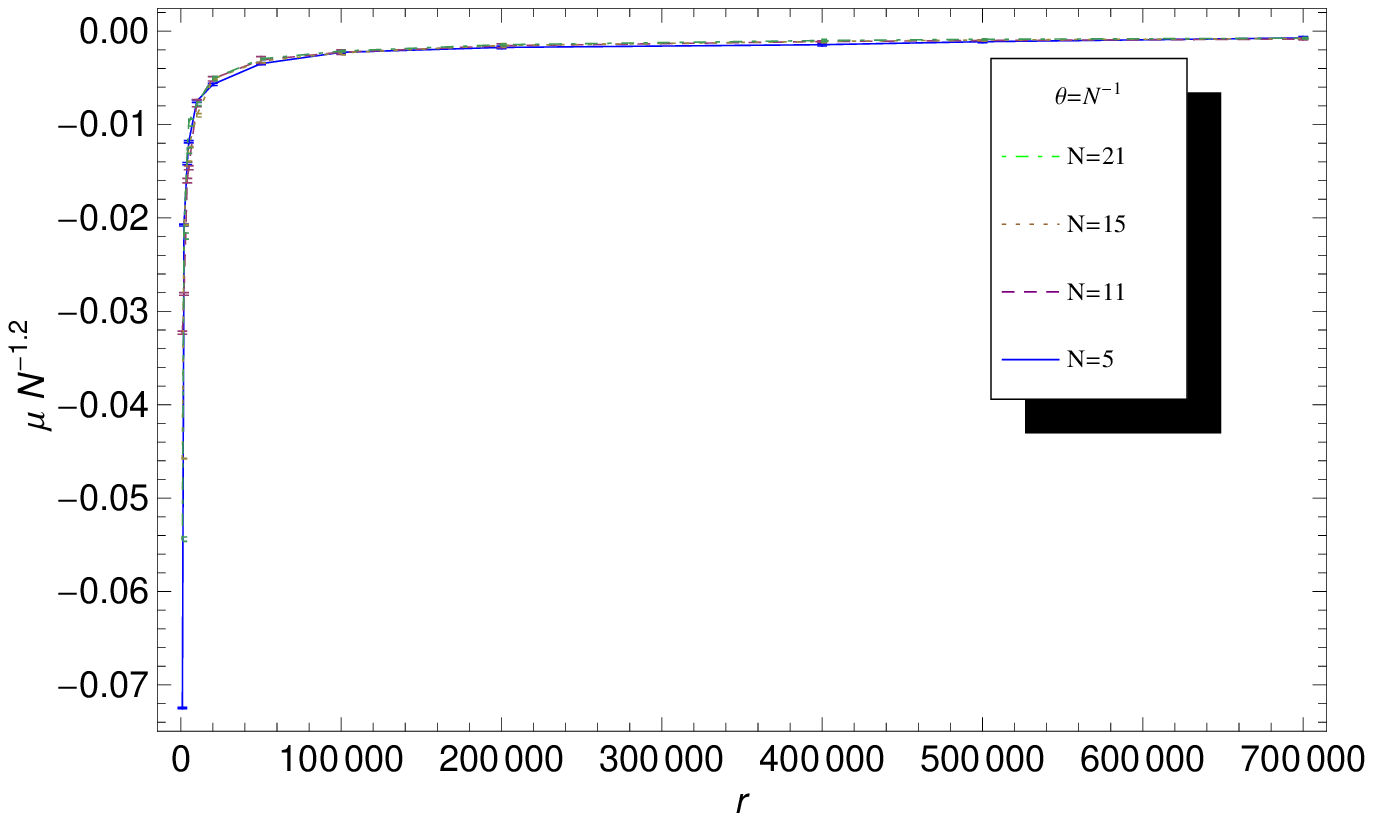}
\includegraphics[scale=0.60]{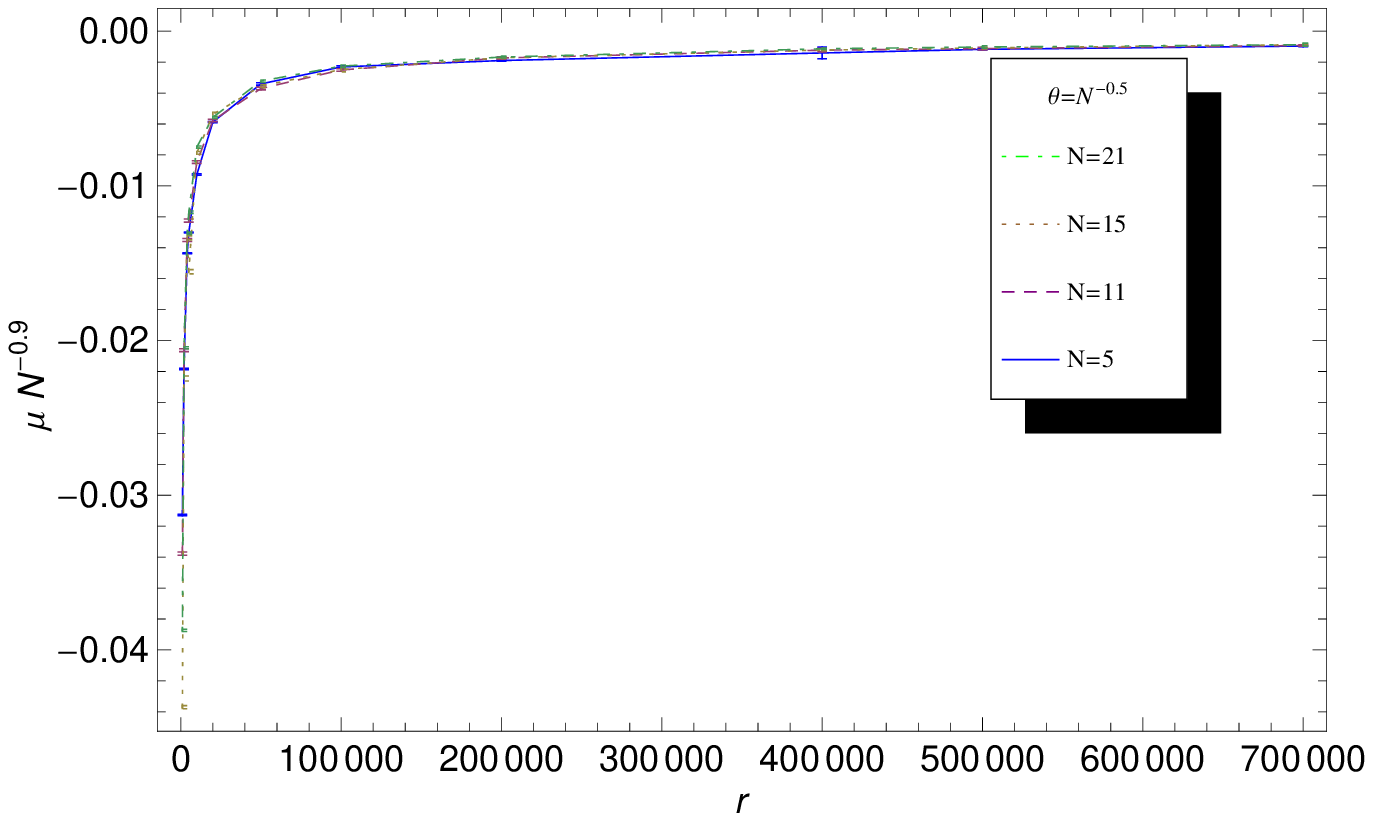}
\end{center}
\vspace{-30pt}
\caption{\sl\footnotesize  Plot of scaled  transition curves  in the different limits  for  $N=5,11,15,21$.
\vspace{-10pt} \normalsize}\label{Figure8}\end{figure}

\begin{figure}[htb]
\vspace{-10pt}
\begin{center}
\includegraphics[scale=0.64]{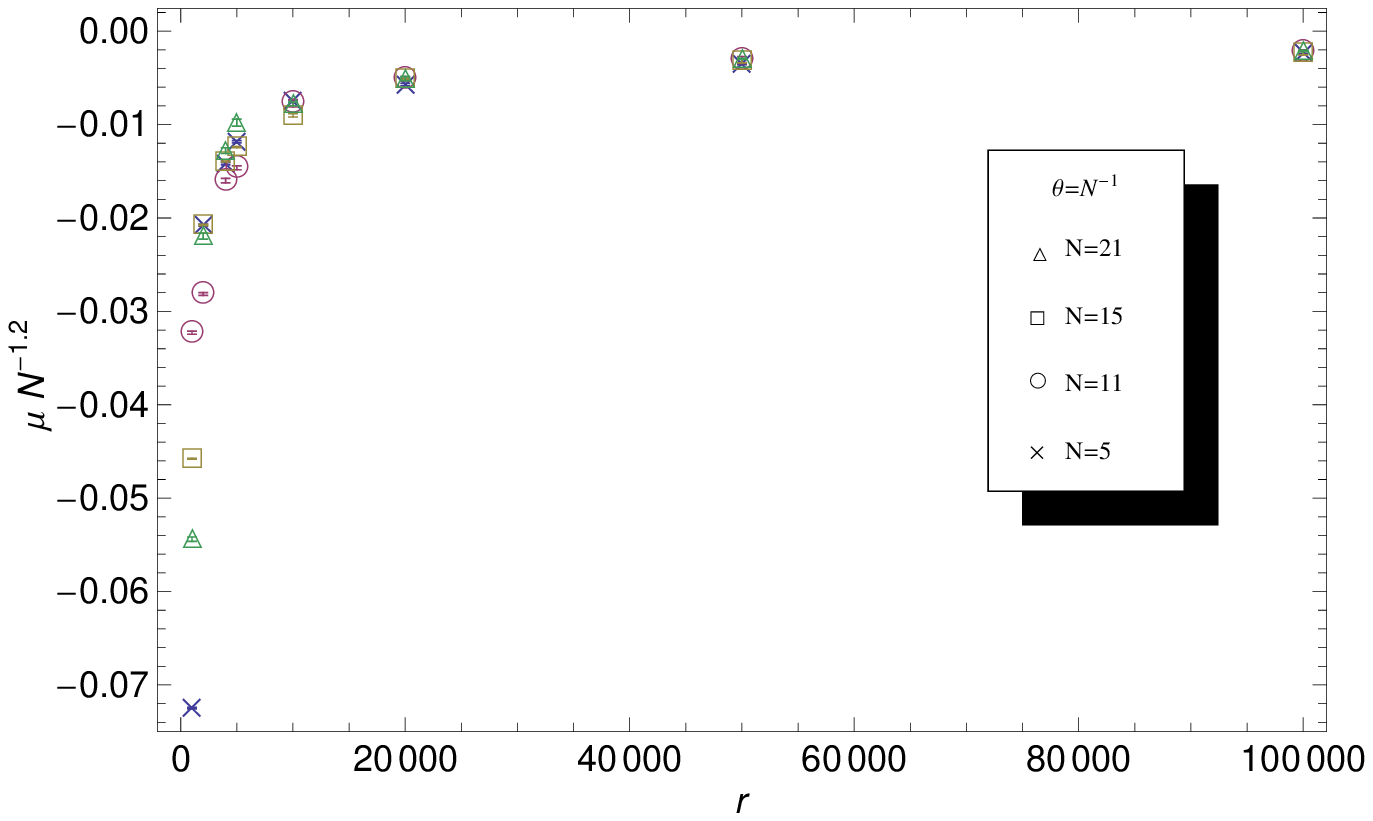}
\includegraphics[scale=0.64]{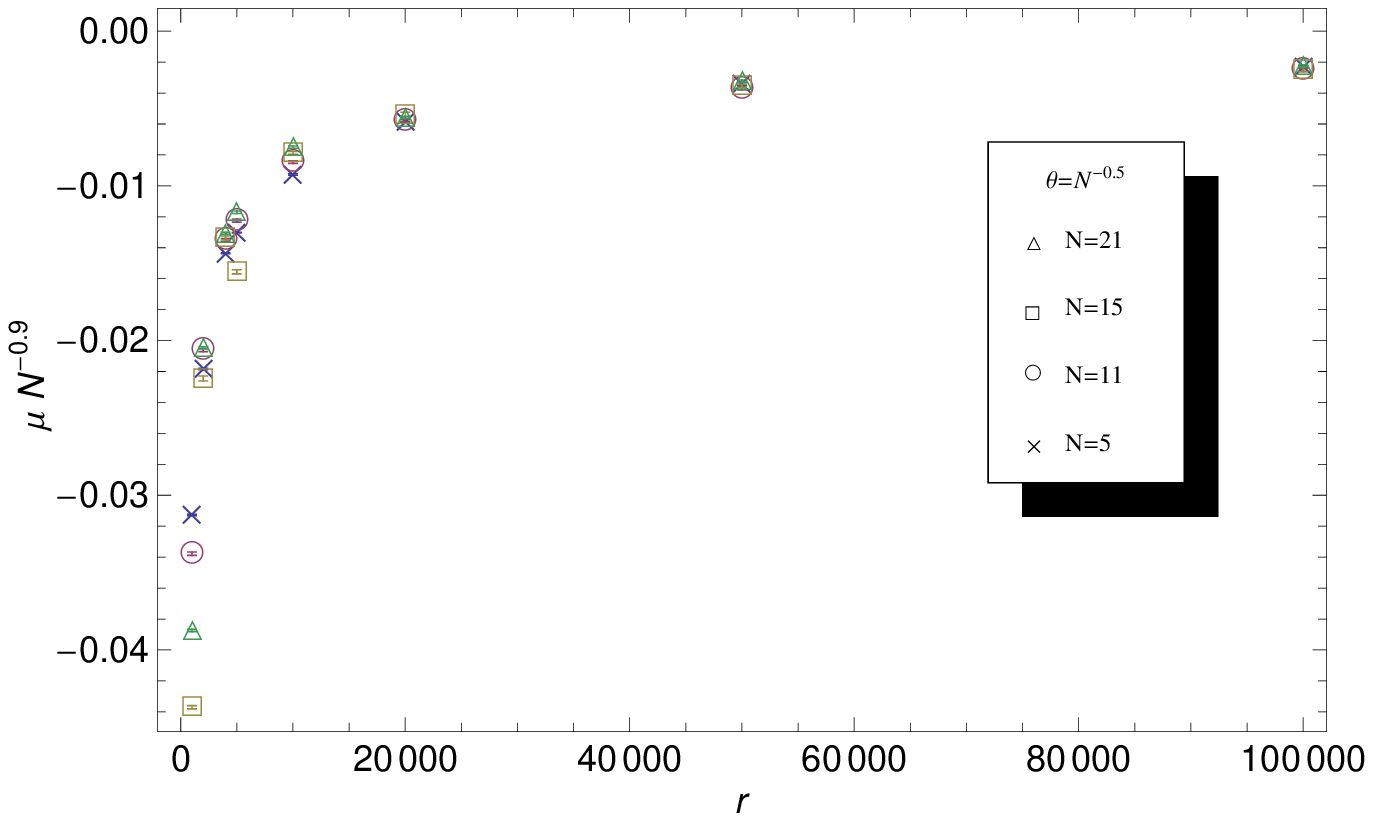}
\end{center}
\vspace{-30pt}\caption{\sl\footnotesize  Plot of scaled  transition curves close to origin for $\theta=N^{-1}, \theta=N^{-0.5} $  and  $N=5,11,15,21.$  \normalsize \vspace{-10pt}
\normalsize}\label{Figure9}\end{figure}

Fig.~\ref{Figure8}  shows the scaled non-uniform-disordered critical lines, each curves scale  in the matrix size with a different scaling factor; for the noncommutative plane limit the curves collapse for a scaling like  for $N^{-0.8}$, for the commutative disc limit the best collapse take place for $N^{-1.2}$ scaling behavior and for the commutative plane for $N^{-0.9}$. Actually  the scalings of of the commutative limits are  not true scalings, in the sense that the collapse takes place for  $r$ large enough, in fact looking closely at the data set Fig.~\ref{Figure9} we notice that the point towards the origin does not collapse. We will discuss this issue later.

The asymptotic behavior for large $r$  can  be  estimated algebraically \cite{Fuzzy-numer-1} from the scaled susceptibility curves and the critical point for the pure potential model, in fact for $N=31$  and $r \to +\infty$
\be
\mu_c (r ) \approx \frac{−17.5}{r}.
\ee
Furthermore using the scaling factors the asymptotic phase transition lines  we have the approximations:
\be
\mu^{\NCP}_c(N,r)\approx \frac{−17.5}{r} N^{-0.8}, \  \  \mu^{\CD}_c(N,r)\approx  \frac{−17.5}{r} N^{-1.2}, \  \ \mu^{\CP}_c(N,r)\approx  \frac{−17.5}{r} N^{-0.9}.
\ee
This large $r$ behavior has an asymptote in zero and fits the scaled critical lines well for $N=31$.

\begin{figure}[htb]
\vspace{-10pt}\begin{center}
\includegraphics[scale=0.64]{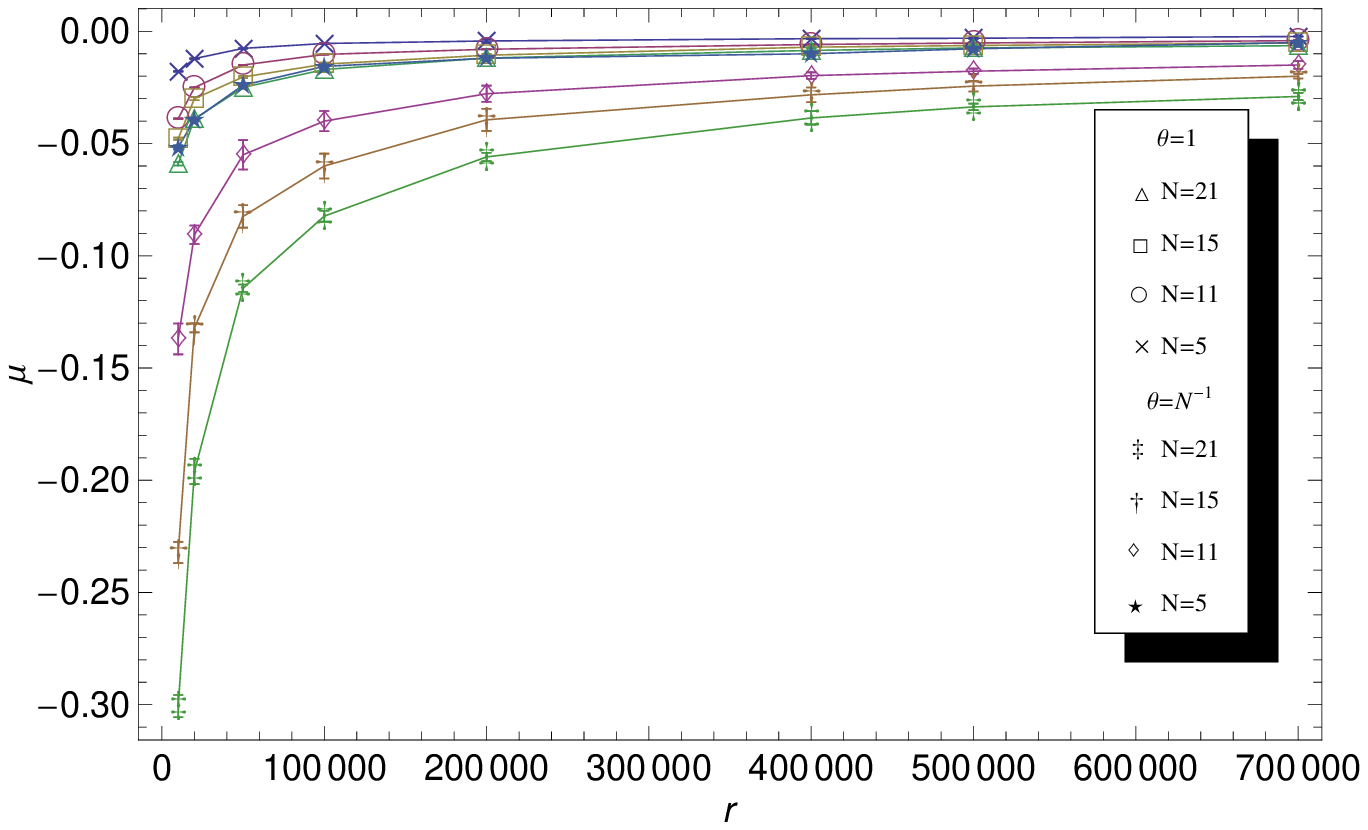}
\includegraphics[scale=0.64]{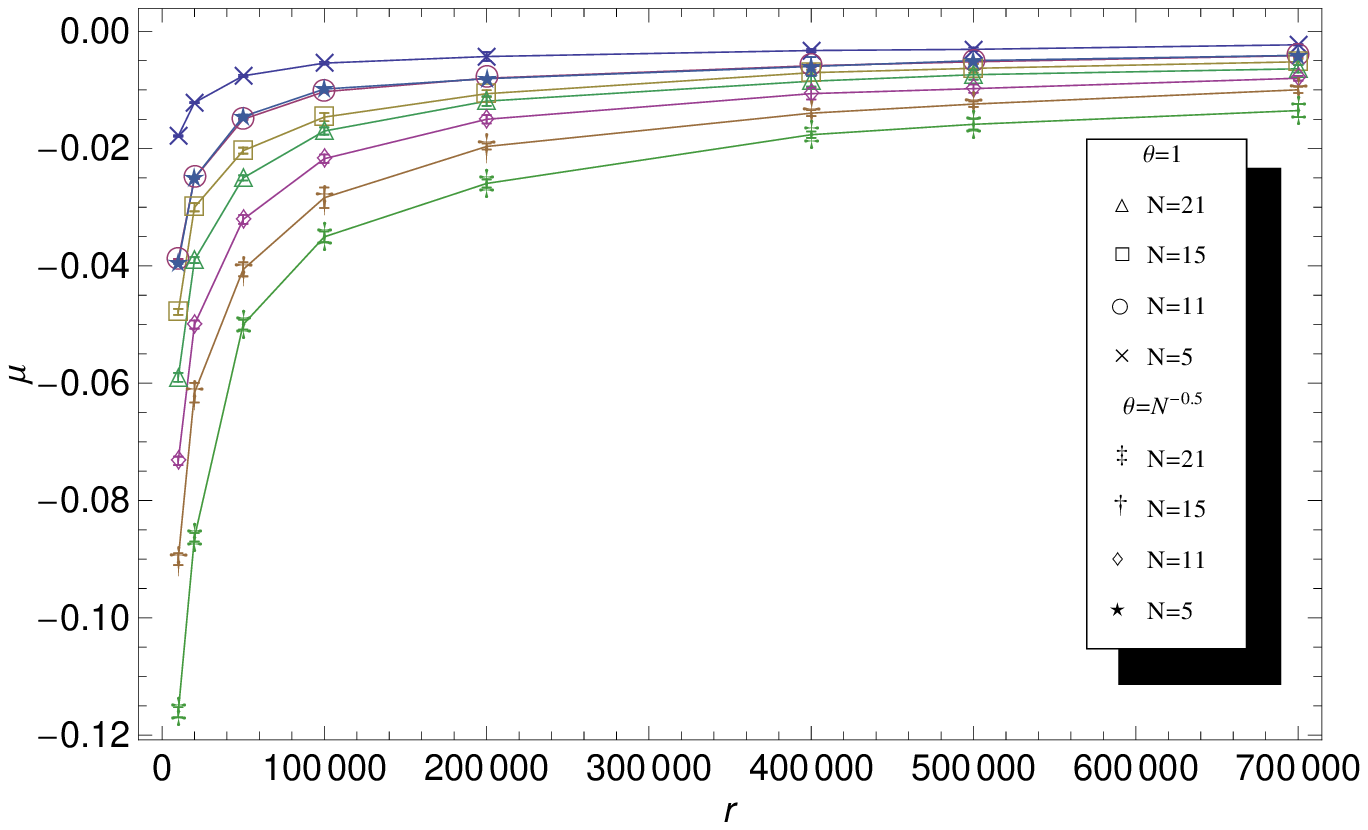}
\end{center}
\vspace{-30pt}\caption{\sl\footnotesize  Plot of the comparisons of transition curves
between the noncommutative plane  limit and commutative disc limit
(left) and   between the noncommutative limit and commutative plane
limit(right) for  $N=5,11,15,21.$\vspace{-10pt} \normalsize}\label{Figure10}\end{figure} 

In  Fig.~\ref{Figure10}  we compare the non scaled
transition curves of the noncommutative limit case and the other
commutative ones, in both cases the difference became bigger for
large $N$ commutative transition curves, the difference is much more
evident  for the commutative disc. The reason of this discrepancy is
the explicit $\theta$ dependence  of the potential term in the
action \eqref{S-Ncom}. The noncommutative limit is characterized by
$\theta=1$  making the model very similar to  the one obtained using
the fuzzy sphere in which the potential part has no dependence on
the noncommutative parameter. Switching to the other to cases we
introduce in the pre factor a dependence on the matrices size,
increasing $N$ for fixed $r$ the effect is to lower the potential
dominance and graphically to lower  the transition curve. For each
$N$ we will find the previous $r_c$  shifted towards lower values
and  since we have for all the cases an finite asymptotic behavior,
in the commutative limits we can lower any point of the transition
curves just increasing $N$.

\section{Conclusions}
\setcounter{equation}{0}
The scalar field action on the fuzzy disc, in analogy with
equivalent simulations on the fuzzy sphere,  shows  the occurrence
of a new nonuniform phase, peculiar of the noncommutative models. We found the existence of three behaviors in the model
described by \eqref{S-Ncom} which arise for all the limit performed.
We studied the scaling properties of the transition curves
disordered-ordered obtaining a scaling coefficient in each case, the
noncommutative case has the same coefficient as the
simulations on the fuzzy sphere  scaling like the square root of the
matrix dimension, the other two cases show different scaling
behavior. Nevertheless looking at the order parameters in all three
cases we can recognize the two phases as the disordered an the
ordered ones.  The critical line for the non-uniform-disordered
phase transition is also found and the noncommutative limit scales
as $N^{-0.8}$ and grows asymptotically  for large  $\mu$ parameters.
In the other two commutative limits  it is possible to consider a good data
collapse using our matrix size approximations only for 
$\mu$ large enough. The asymptotic behaviors of are well approximated for large
$r$ by a pure potential model. We have confirmed  that  even for
this models that the non-uniform phase is  connected to the
dominance  of the minimized potential over the kinetic term via the
phase space volume of the largest orbit. The geometry for a
scalar field theory,  defined on  a fuzzy space, is  defined by  the
Laplacian operator   which correspond to a particular kinetic
term of the action, thus the new phase should  be  insensitive  on
the underlying  geometry. In particular, this argument justifies the
similar behavior of the noncommutative limit comparing  to the
results found for the fuzzy sphere studied in  \cite{Fuzzy-numer-1,Fuzzy-numer-2,panero-2}. 
A substantial difference from the fuzzy sphere case  is in the
commutative limits, in fact the $\theta$ dependence of the potential
part in each two cases modify  the data collapse of the transition
curves disordered non-uniform  making a good collapse possible only
for large $r$ respect $N$. In the fuzzy sphere models the usual
scalar field action cannot be used as an
approximation of the scalar field theory on the commutative sphere
due the persistence of the non ordered phase in the continuous
limit. The problem is patched with an additional term used to
increase the weight of the kinetic part or equivalently to decrease
the influence of the potential term in order to suppress the
anomalous phase \cite{Fuzzy-Action-mod}.   In our model, considering the commuting limits, the
action of the $\theta$ on the potential term act naturally  reducing
the ratio between the potential part and the kinetic one, so we can
surmise  that in the continuous limits the non-uniform phase does
not exist. The ability to differentiate the commutative and noncommutative cases is one of the main advantages of the method described in this paper.

\paragraph{Acknowledgments}
F.~Lizzi acknowledges
support by CUR Generalitat de Catalunya under project FPA2010-20807
and the {\sl Faro} project {\sl Algebre di Hopf, differenziali e di
vertice in geometria, topologia e teorie di campo classiche e
quantistiche} of the Universit\`a di Napoli {\sl Federico II}.


\begin{thebibliography}{99}

\bibitem{Szaboreview} R.~J.~Szabo,
  ``Quantum field theory on noncommutative spaces,''
  Phys.\ Rept.\  {\bf 378} (2003) 207
  [hep-th/0109162].
  %%CITATION = HEP-TH/0109162;%%

\bibitem{Connesbook} A. Connes, \textit{Noncommutative Geometry},
    Academic Press, 1984.

\bibitem{Landi} G. Landi, {\it  An Introduction to
    Noncommutative Spaces and their Geometries}, {\sl Springer
    Lecture Notes in Physics 51}, Springer Verlag (Berlin
    Heidelberg) 1997. arXiv:hep-th/9701078.
  %%CITATION = HEP-TH/9701078;%%

\bibitem{Ticos} J.M.~Gracia-Bondia, J.C.~Varilly, H.~Figueroa, {\it
    Elements of Noncommutative Geometry}, Birkhauser, 2000.


\bibitem{Fuzzy-s} J.~Madore,
  ``The Fuzzy sphere,''
  Class.\ Quant.\ Grav.\  {\bf 9} (1992) 69.
  %%CITATION = CQGRD,9,69;%%

\bibitem{BalSeckinSachin} A.~P.~Balachandran, S.~Kurkcuoglu and
    S.~Vaidya,
  ``Lectures on fuzzy and fuzzy SUSY physics,''
  Singapore, Singapore: World Scientific (2007) 191 p.
  [hep-th/0511114].
  %%CITATION = HEP-TH/0511114;%%

\bibitem{Medina} J.~Medina,
  ``Fuzzy Scalar Field Theories: Numerical and Analytical Investigations,''
  arXiv:0801.1284 [hep-th].
  %%CITATION = ARXIV:0801.1284;%%

\bibitem{fuzzy-numer-3}J.~Medina, W.~Bietenholz, F.~Hofheinz and D.~O'Connor,
  %``Field theory simulations on a fuzzy sphere: An Alternative to the lattice,''
  PoS LAT {\bf 2005} (2006) 263
  [hep-lat/0509162].
  %%CITATION = HEP-LAT/0509162;%%

\bibitem{Gronewold} H. Gr\"onewold, ``On the Principles of Quantum
    Mechanics'', Physica {\bf12} (1946)~405.
%%CITATION = PHYSA,12,405;%%



\bibitem{Moyal} J.~E.~Moyal,
  ``Quantum Mechanics as a Statistical Theory,''
  Proc.\ Cambridge Phil.\ Soc.\  {\bf 45} (1949)~99.
%%CITATION = PCPSA,45,99;%%

\bibitem{GLV08} S.~Galluccio, F.~Lizzi and P.~Vitale,
  ``Twisted Noncommutative Field Theory with the Wick-Voros and Moyal Products,''
  Phys.\ Rev.\ D {\bf 78} (2008) 085007
  [arXiv:0810.2095 [hep-th]].
  %%CITATION = AR

\bibitem{montecarlo-3} J. Zinn-Justin, {\it Quantum Field Theory and Critical Phenomena}, Oxford University Press, Second edition (1993).

\bibitem{montecarlo-4} I. Montvay and G. M\"unster, {\it Quantum Field Theory on a Lattice}, Cambridge University Press (1997).



\bibitem{Fuzzy-susy}
B.~Ydri,
  ``Impact of Supersymmetry on Emergent Geometry in Yang-Mills Matrix Models II,''
  Int.\ J.\ Mod.\ Phys.\ A {\bf 27} (2012) 1250088
  [arXiv:1206.6375 [hep-th]].
  %%CITATION = ARXIV:1206.6375;%%

\bibitem{Fuzzy-YM}
 R.~Delgadillo-Blando, D.~O'Connor and B.~Ydri,
  ``Geometry in Transition: A Model of Emergent Geometry,''
  Phys.\ Rev.\ Lett.\  {\bf 100} (2008) 201601
  [arXiv:0712.3011 [hep-th]].
  %%CITATION = ARXIV:0712.3011;%%
  
  \bibitem{Num-WG1}
B.~Spisso, R.~Wulkenhaar,``A numerical approach to harmonic non-commutative spectral field theory,''
  Int.\ J.\ Mod.\ Phys.\ A {\bf 27} (2012) 1250075
  [arXiv:1111.3050v4 [math-ph]].
  %%CITATION = ARXIV:1111.3050;%%
  
\bibitem{Num-WG2}
B.~Spisso, ``A numerical approach to harmonic non-commutative spectral field theory''
  [arXiv:1111.2871v1 [hep-th]].
%%CITATION = ARXIV:1111.2951;%%


\bibitem{fuzzy1}
  F.~Lizzi, P.~Vitale and A.~Zampini,
  ``The Fuzzy disc,''
  JHEP {\bf 0308} (2003) 057
  [hep-th/0306247].
  %%CITATION = HEP-TH/0306247;%%

\bibitem{fuzzybal} A.~P.~Balachandran, K.~S.~Gupta and S.~Kurkcuoglu,
  ``Edge currents in noncommutative Chern-Simons theory from a new matrix model,''
  JHEP {\bf 0309} (2003) 007
  [hep-th/0306255].
  %%CITATION = HEP-TH/0306255;%%


\bibitem{fuzzy2}
  F.~Lizzi, P.~Vitale and A.~Zampini,
  ``The Beat of a fuzzy drum: Fuzzy Bessel functions for the disc,''
  JHEP {\bf 0509} (2005) 080
  [hep-th/0506008].
  %%CITATION = HEP-TH/0506008;%%

\bibitem{fuzzy3}
  F.~Lizzi, P.~Vitale and A.~Zampini,
  ``From the fuzzy disc to edge currents in Chern-Simons theory,''
  Mod.\ Phys.\ Lett.\ A {\bf 18} (2003) 2381
  [hep-th/0309128].
  %%CITATION = HEP-TH/0309128;%%

\bibitem{fuzzy4}
  F.~Lizzi, P.~Vitale and A.~Zampini,
  ``The fuzzy disc: A review,''
  J.\ Phys.\ Conf.\ Ser.\  {\bf 53} (2006) 830.
  %%CITATION = 00462,53,830;%%

\bibitem{Zampthesis}
A.~Zampini,
  ``Applications of the Weyl-Wigner formalism to noncommutative geometry,''
  hep-th/0505271.
  %%CITATION = HEP-TH/0505271;%%

\bibitem{KobayashiAsakawa}  S.~Kobayashi and T.~Asakawa,
  ``Angles in Fuzzy Disc and Angular Noncommutative Solitons,''
  arXiv:1206.6602 [hep-th].
  %%CITATION = ARXIV:1206.6602;%%

\bibitem{UV-IR}
S.~Minwalla, M.~Van Raamsdonk and N.~Seiberg,
  ``Noncommutative perturbative dynamics,''
  JHEP {\bf 0002} (2000) 020
  [hep-th/9912072].
  %%CITATION = HEP-TH/9912072;%%

\bibitem{UV-IR-1} M.~Chaichian, A.~Demichev and P.~Presnajder,
  ``Quantum field theory on noncommutative space-times and the persistence of ultraviolet divergences,''
  Nucl.\ Phys.\ B {\bf 567} (2000) 360
  [hep-th/9812180].
  %%CITATION = HEP-TH/9812180;%%
  
 \bibitem{montecarlo-1} M. E. Newman and G. T. Barkema, {\it Monte Carlo Methods in Statistical Physics }, Oxford University Press (2002).

\bibitem{montecarlo-2} W. Janke, ``Statistical Analysis of Simulations: Data Correlations and Error 
Estimation'', published in Quantum Simulations of Complex Many-Body Systems:
From Theory to Algorithms, Lecture Notes John von Neumann Institute for Computing, 
J\"ulich,NIC Series, Vol. 10, ISBN 3-00-009057-6 (2002) 423-445.

\bibitem{Fuzzy-numer-1}
X.~Martin,
  ``A Matrix phase for the $\varphi^4$ scalar field on the fuzzy sphere,''
  JHEP {\bf 0404} (2004) 077
  [hep-th/0402230].
  %%CITATION = HEP-TH/0402230;%%
  
  \bibitem{panero-1} 
  M.~Panero,
  ``Numerical simulations of a non-commutative theory: The Scalar model on the fuzzy sphere,''
  JHEP {\bf 0705} (2007) 082
  [hep-th/0608202].
  %%CITATION = HEP-TH/0608202;%%

\bibitem{Fuzzy-numer-2} F.~Garcia Flores, X.~Martin and D.~O'Connor,
  ``Simulation of a scalar field on a fuzzy sphere,''
  Int.\ J.\ Mod.\ Phys.\ A {\bf 24} (2009) 3917
  [arXiv:0903.1986 [hep-lat]].
  %%CITATION = ARXIV:0903.1986;%%


  
 \bibitem{Fuzzy-numer-3}
 T.R.~Govindarajan, S.~Digal, K.S.~Gupta, X.~Martin, ``Phase structures in fuzzy geometries'', 	      [arXiv:1204.6165v1 [hep-th]]
 
 
 \bibitem{Fuzzy-numer-4}
 C.R.~Das, S.~Digal, T.R.~Govindarajan, ``Finite temperature phase transition of a single scalar field on a fuzzy sphere'', 	[arXiv:0706.0695v2 [hep-th]]
 
 
\bibitem{panero-2}
M.~Panero,
  ``Quantum field theory in a non-commutative space: Theoretical predictions and numerical results on the fuzzy sphere,''
  SIGMA {\bf 2} (2006) 081
  [hep-th/0609205].
  %%CITATION = HEP-TH/0609205;%%
 
 \bibitem{Fuzzy-Action-mod}
B. P. Dolan, D. O'Connor and P. Pre\v{s}najder, ``Matrix $\varphi^4$ models on the fuzzy
sphere and their continuum limits'', JHEP {\bf 03} (2002) 013 [hep-th/0109084].
  %%CITATION = HEP-TH/0109084;%%
 
 \bibitem{Fuzzy-numer-5}
 C.R.~Das, S.~Digal, T.R.~Govindarajan, ``Spontaneous symmetry breakdown in fuzzy spheres'',
  Mod.Phys.Lett.A24 	[arXiv:0801.4479v2 [hep-th]]
 
 \bibitem{Fuzzy-numer-6}
  S.~Digal, T.R.~Govindarajan ``Topological stability of broken symmetry on fuzzy spheres''
  Mod. Phys. Lett. A, Vol. 27, No. 14 [arXiv:1108.3320v2 [hep-th]]
  
  \bibitem{striped}
   S.S.~Gubser and S.L.~Sondhi, ``Phase structure of non-commutative scalar field theories'' 
   Nucl. Phys. \textbf{B605}, 395 (2001) [hep-th/0006119].

\bibitem{striped-1}
  J.~Ambjorn, S.~Catterall, ``Stripes from (noncommutative) stars'' 
  Phys. Lett. \textbf{B549}  [arXiv:hep-lat/0209106v3].
  
  
  


\end{thebibliography}
\end{document}